\documentclass[]{jfm}

\usepackage{graphicx}
\usepackage{newtxtext}
\usepackage{newtxmath}
\usepackage{natbib}
\usepackage{threeparttable}
\usepackage{subfigure}
\usepackage{mleftright}
\usepackage{amssymb,bbding,oplotsymbl,MnSymbol}
\usepackage{relsize}

\usepackage{hyperref}
\hypersetup{
    colorlinks = true,
    urlcolor   = blue,
    citecolor  = orange,
}

\definecolor{maroon}{rgb}{0.64, 0.08, 0.18}	

\definecolor{d_cyan}{rgb}{0.0, 0.45,0.74}

\def\spopa#1#2{{\frac{d#1}{d#2}}}

\def\spopasq#1#2{{{d^2#1}\over{d#2^2}}}

\newcommand{\RomanNumeralCaps}[1]

\usepackage{amsmath,xcolor}
 
\title{An experimental study of flow near an advancing contact line: a rigorous test of theoretical models}

\author{Charul Gupta\aff{1},
Anjishnu Choudhury\aff{2},
  Lakshmana D Chandrala\aff{1},
 \and Harish N Dixit\aff{1,3}
  \corresp{\email{hdixit@mae.iith.ac.in}},
 }

\affiliation{\aff{1}Department of Mechanical \& Aerospace Engineering, Indian Institute of Technology Hyderabad, India
\aff{2} PMMH, CNRS, ESPCI Paris, Université PSL, Sorbonne Université, Université de Paris, F-75005, Paris, France
\aff{3}Polymers \& Biosystems Engineering, Center for Interdisciplinary Programs, Indian Institute of Technology Hyderabad, India}

\begin{document}
\maketitle


\begin{abstract}
The flow near a moving contact line depends on the dynamic contact angle, viscosity ratio, and capillary number. We report experiments involving immersing a plate into a liquid bath, concurrently measuring the interface shape, interfacial velocity, and fluid flow using digital image processing and particle image velocimetry. All experiments were performed at low plate speeds to maintain small Reynolds and capillary numbers for comparison with viscous theories. The dynamic contact angle, measured in the viscous phase, was kept below $90^{\circ}$, an unexplored region of parameter space. An important aim of the present study is to provide valuable experimental data using which new contact line models can be developed and validated. 
Interface shapes reveal that the strong viscous bending predicted by theoretical models is absent in the experimental data. The flow field is directly compared against the prediction from the viscous theory of \cite{huh1971hydrodynamic} but with a slight modification involving the curved interface. Remarkable agreement is found between experiments and theory across a wide parameter range. The prediction for interfacial speed from \cite{huh1971hydrodynamic} is also in excellent agreement with experiments except in the vicinity of the contact line. Material points along the interface were found to rapidly slow down near the contact line, thus alleviating the singularity at the moving contact line. To the best of our knowledge, such a detailed test of theoretical models has not been performed before and we hope the present study will spur new modeling efforts in the field.
\end{abstract}

\begin{keywords}
Authors should not enter keywords on the manuscript, as these must be chosen by the author during the online submission process and will then be added during the typesetting process (see \href{https://www.cambridge.org/core/journals/journal-of-fluid-mechanics/information/list-of-keywords}{Keyword PDF} for the full list).  Other classifications will be added at the same time.
\end{keywords}

{\bf MSC Codes }  {\it(Optional)} Please enter your MSC Codes here

\section{\label{sec:Intro}INTRODUCTION}
One of the most ubiquitous fluid flow phenomena in fluid mechanics involves the motion of two immiscible fluids on a solid surface. Examples range from the sliding of water drops on windowpanes, the spreading of droplets on paper in ink-jet printing to the coating industry where a thin film of a liquid is deposited on a solid surface. A common theme in all these problems is the presence of a three-phase contact line at the intersection of the three phases (typically solid, liquid, and gas). The value of the contact angle, often measured in the liquid phase, determines the wettability of the surface. In static systems on smooth substrates, the value of the contact angle is related to the surface energies of the three surfaces and is given by Young's law. But when the contact line is in motion, the problem becomes significantly more complex. First, the problem ceases to be in thermodynamic equilibrium, thus the dynamic contact angle, $\theta_d$, deviates from the static angle, $\theta_s$. Second, application of the standard no-slip boundary condition at the solid surface leads to a stress singularity at the contact line \citep{huh1971hydrodynamic} which leads to a logarithmic divergence in the dissipation as one approaches the contact line (\cite{bonn2009wetting}). Third, the interface shape departs from the static shape and needs to be determined simultaneously along with the flow field. Surprisingly, in spite of the singularity, \cite{huh1971hydrodynamic} (HS71 hereafter) show that the flow fields away from the contact line are well defined and depend only on the contact angle of the wedge flow and the viscosity ratio. 

\begin{figure}
\centering
  \includegraphics[trim = 0mm 0mm 0mm 0mm, clip, angle=0,width=0.4\textwidth]{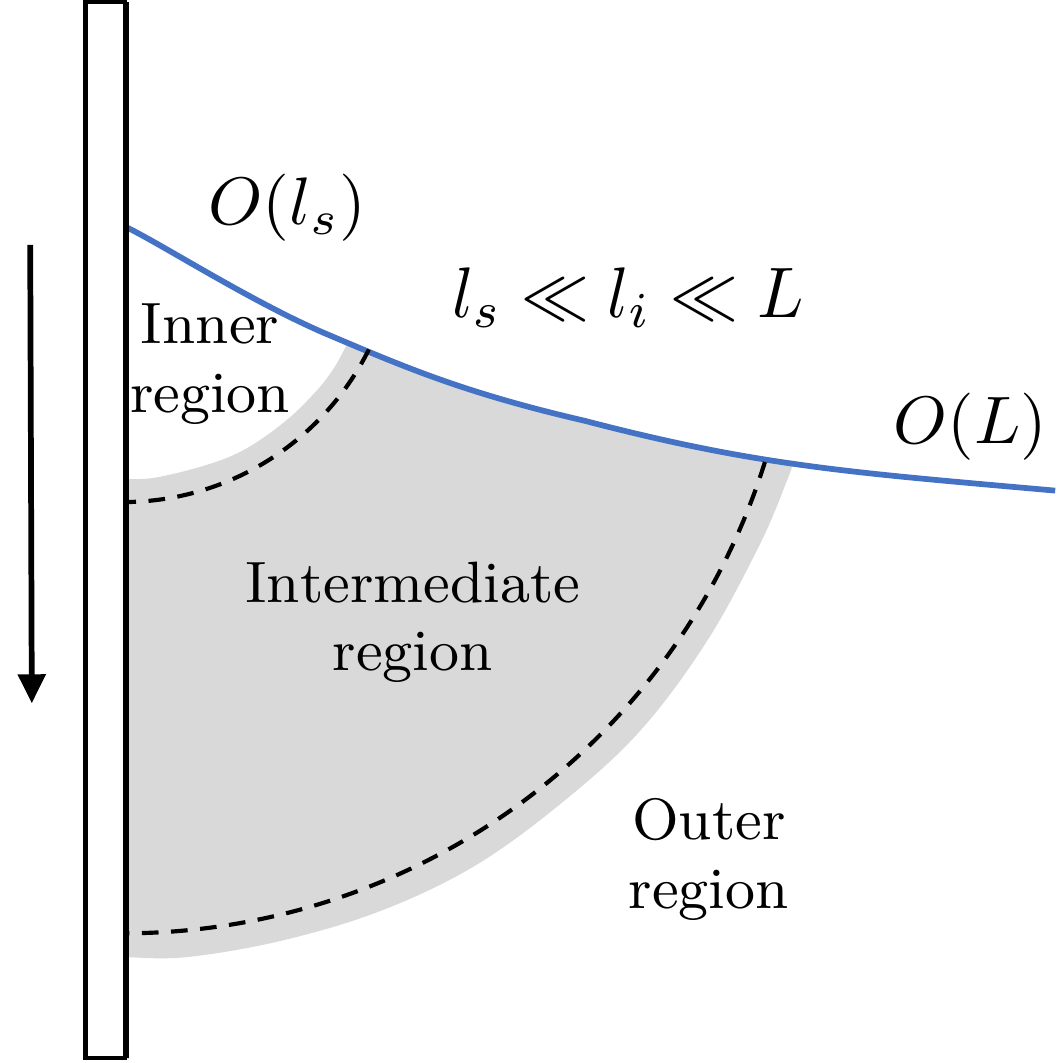}
\caption{Different regions of the flow near a moving contact line as per \cite{cox1986dynamics}}
\label{fig:Cox_matching_regions}
\end{figure}

Over the last few decades, a number of important advances have been made to deal with the above issues in the viscous limit, i.e. where the inertia of the fluid is negligible. Such a limit is obtained either by using very viscous fluids or by restricting the domain of interest to regions very close to the contact line such that the local Reynolds number (ratio of inertia to viscous forces) remains small. These viscous theories (\cite{blake1969kinetics,de1985wetting,cox1986dynamics,shikhmurzaev1993moving}), derived in the limit $Re \ll 1$ and $Ca \ll 1$, aim to relate the dynamic contact angle, $\theta_d$, to the contact line velocity. The singularity at the moving contact line is relieved by incorporating additional physics at the moving contact line. In particular, \cite{cox1986dynamics} introduced slip at the contact line, divided the flow into three regions as shown schematically in figure \ref{fig:Cox_matching_regions} and used matched asymptotic techniques to match the `inner' slip-dominated region to a geometry dependent `outer' region using an `intermediate' region. This results in a simple model for the dynamic contact angle, $\theta_d$ as a function of the capillary number, $Ca=\mu_{\text{oil}}U/\gamma$, the viscosity ratio, $\lambda=\mu_{\text{air}}/\mu_{\text{oil}}$, and the ratio of the slip-length to a characteristic scale of the outer region, $\epsilon = l_s/L$. For drops sliding down an incline, the characteristic outer scale, $L$, is a typical size of the drop whereas for plate advancing experiments, $L$ could be taken to be the capillary length. More recent models, such as the interface formation model of \cite{shikhmurzaev1997moving}, are similar to Cox's model in one key aspect. In all these models, the flow field in the intermediate region is identical to the HS71's solution. Hence, a careful test of the HS71's solution will also serve to test the intermediate region in Cox's model and several other similar models. Using careful experiments, it should be possible to determine the nature of the flow at a length scale much smaller than the scale of the outer region which is likely to correspond to the intermediate region shown in figure \ref{fig:Cox_matching_regions}.

A number of simple theoretical slip models have been proposed to alleviate the singularity at the moving contact line. \cite{dussan1976moving} explored various models for slip along the moving plate and showed that the flow fields are the same and independent of the exact nature of the slip model when viewed at the `meniscus' length scale, i.e. at a length scale far away from the slip length scale. \cite{sheng1992immiscible} employed three phenomenological slipping models and determined the flow using numerical techniques. In recent years, \cite{kirkinis2013hydrodynamic,kirkinis2014moffatt} employed an algebraic slip model with perfect slip at the contact line and no-slip at a finite distance away from the contact line and obtained the flow fields in the `inner region' of the flow. This approach was further refined and extended by \cite{febres2017existence} for a two-fluid system. In the above two approaches, an extra complex parameter, $n$, is introduced in the streamfunction similar to what was employed by \cite{moffatt1964viscous}, and $n$ is determined by solving an eigenvalue problem. Unfortunately, there appear to be infinitely many possibilities for $n$ each of which results in a different flow field making it difficult to directly test the results against experiments. Theoretical models have been developed for the dynamic interface shape, for example, \cite{dussan1991} extended Cox's model by incorporating the static shape in the outer region. A more direct approach was used by \cite{chan2013hydrodynamics} and \cite{chan2020cox} who derived a differential equation for the dynamic meniscus incorporating slip at the moving wall. \cite{kulkarni2023stream} recently obtained theoretical expressions for the streamfunction using a variety of new slip boundary conditions. Some of the above theoretical models are discussed in detail in \S\ref{sec:theory}.


While a great number of experimental studies have focused on testing the relationship between the dynamic contact angle and the capillary number, there are relatively few studies focusing on the flow fields and the interfacial speeds. In an early experimental study, \cite{dussan1974motion} studied the motion of highly viscous liquid drops (honey and glycerol, $\lambda \ll 1$) and showed that the large-scale motion in the vicinity of the advancing contact line is of the `rolling-type', i.e., the fluid particles at the interface approach the contact line and then roll beneath the drop. \cite{hoffman1975study} carried out systematic measurements of the dynamic contact angle by studying motion in a horizontal capillary tube with five different fluids and showed the existence of a universal relationship between the dynamic contact angle and the capillary number. \cite{dussan1991} carried out controlled dipping of a tube inside a liquid bath and measured the shape of the interface. They amended the model of Cox by incorporating the interface shape from the outer solution allowing them to theoretically match the interface shape with experiments over a wide range of length scales. \cite{le2005shape} carried out experiments with silicone oil and PDMS drops and focused on the shape of the drop moving down an incline. They also compared the dynamic contact angles with various models published in the literature. All the above studies focus on the variation of dynamic contact angle with speed. \cite{chen1997velocity} carried out tube-advancing experiments in highly viscous liquids (PDMS) and obtained flow fields using PIV techniques. All their experiments were carried out at moderate to high $Ca$ and at obtuse angles, i.e. $\theta_d > 90^{\circ}$. To compare the flow fields with the fixed wedge theory of HS71's theory, \cite{chen1997velocity} allowed the angle to vary along the interface and incorporated this angle in the solution of HS71's theory. It has to be noted that this solution, referred to as the `modulated wedge solution', is not an exact solution of the biharmonic equation in a curved wedge. One common approach to alleviate the singularity in HS71's theory is by introducing slip at the moving contact line. 

\begin{figure}
\centering
\includegraphics[trim = 0mm 0mm 0mm 0mm, clip, angle=0,width=0.7\textwidth]{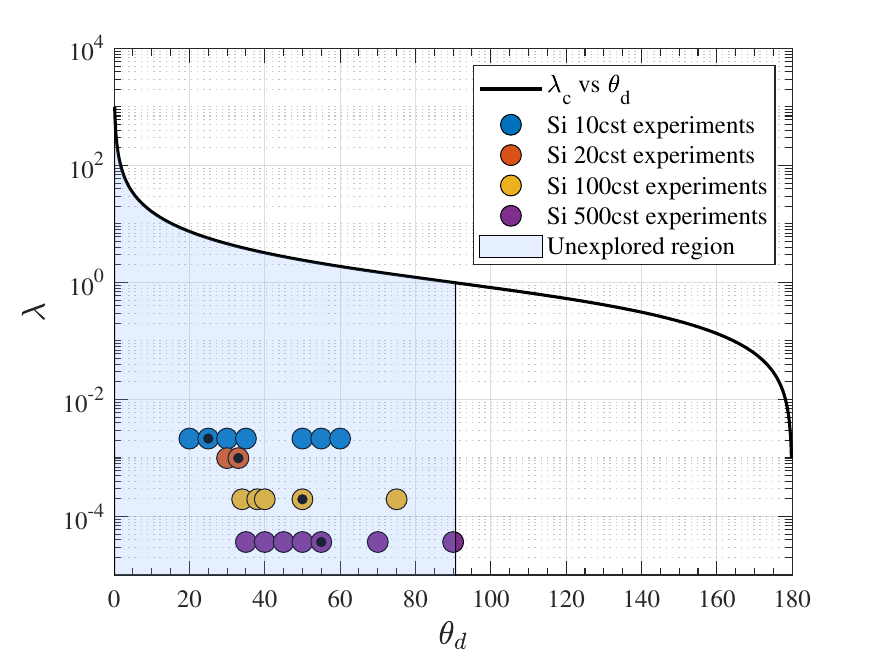}
\caption{Operating regime map in the viscosity ratio $\lambda$ and dynamic contact angle $\theta_d$ plane. The solid black curve represents the theoretical prediction of \cite{huh1971hydrodynamic} for critical viscosity ratio, $\lambda_c$, corresponding to a motionless interface. All symbols correspond to experimental data points for fluids with different viscosities. Streamfunction plots will be shown later in the paper for data points shown with a black dot. The shaded region shows unexplored parameter space in the literature for advancing contact lines.}
\label{fig:operating_regime_map}
\end{figure}

To the best of our knowledge, there are no systematic experiments at low $Re$ and low $Ca$ with $\theta_d < 90^{\circ}$ where direct comparison with theory has been carried out.
This regime is usually difficult to achieve in experiments with advancing drops since drops usually assume a contact angle greater than $90^{\circ}$ before moving down an incline. To overcome this difficulty, we use plate-advancing experiments with a glass plate dipped into a bath of silicone oil at controlled speeds. As per the theoretical framework of HS71, the problem is fully determined by just two parameters: the viscosity ratio, $\lambda$, and the dynamic contact angle, $\theta_d$. Figure \ref{fig:operating_regime_map} shows the parameter regime explored in the present work and the shaded region corresponds to the region of parameter space where the dynamic contact angle is acute. The solid curve in figure \ref{fig:operating_regime_map} indicates a critical viscosity ratio from HS71 theory where the interface remains motionless. As per the theory of HS71, in the shaded region, all fluid particles at the interface approach an advancing moving contact line. This has been shown to be true in several studies for advancing contact lines, except that all the earlier studies are for $\theta_d > 90^{\circ}$. The present study fills this gap in the literature by conducting systematic experiments for advancing contact lines with $\theta_d<90^{\circ}$, thus providing valuable data against which numerical models can be tested. Further, the interfacial velocity from the present study can also be directly used as boundary conditions in numerical models to alleviate the singularity at the dynamic contact line.

The paper is organized as follows. In \S\ref{sec:expt_setup}, we describe the experimental setup and flow visualization techniques employed. A review of earlier theoretical work is discussed in \S\ref{sec:theory} along with the derivation of the modulated wedge solution suitable for the present geometry. Key results, which include flow fields obtained from PIV experiments, determining the interface shape and interface speeds, and comparison with theoretical predictions is discussed in \S\ref{sec:results}. We conclude the paper in \S\ref{sec:summary_discussion} with a brief discussion of key outcomes and future directions.

\section{Experimental setup and data analysis}\label{sec:expt_setup}
\begin{figure}
\centering
\includegraphics[trim = 0mm 0mm 0mm 0mm, clip, angle=0,width=0.9\textwidth]{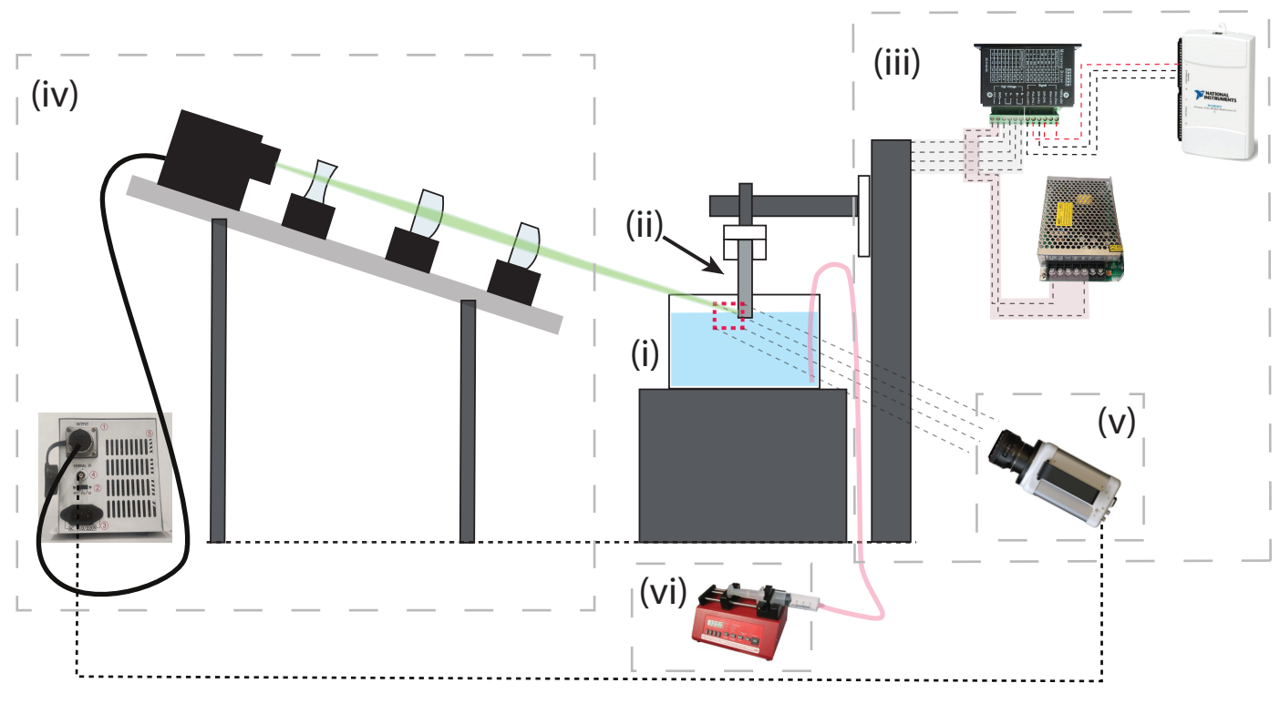}
\caption{A schematic of the PIV experimental setup with the following key components: (i) rectangular tank, (ii) glass slide, (iii) motorized traverse mechanism with DC power source and controller, (iv) laser with controller and power source along with associated optics to produce a thin laser sheet, (v) camera with a macro lens, (vi) programmable syringe pump used to maintain a constant liquid level.}
\label{fig:schematic_setup}
\end{figure}

\begin{figure}
\centering
\includegraphics[trim = 0mm 0mm 0mm 0mm, clip, angle=0,width=0.55\textwidth]{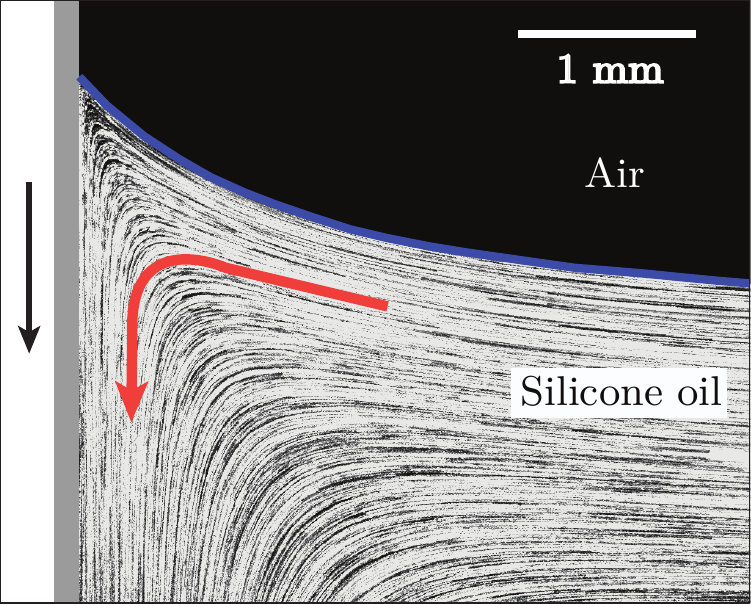}
\caption{A streakline image for 500 cSt Silicone oil at $Re = 3.92 \times 10^{-4}$ and $Ca = 4.19 \times 10^{-3}$. The glass plate, represented by a gray vertical slab is immersed in a liquid bath at constant speed. The blue solid curve represents the interface between the liquid and air, whereas the red arrow represents the direction of the flow in the liquid phase.}
\label{fig:streaklines-plot}
\end{figure}

The experimental setup used in the current study is shown schematically in figure \ref{fig:schematic_setup}. A thin glass plate with dimensions $75\text{mm}\times 25\text{mm} \times 1\text{mm}$ was dipped into an acrylic tank of dimensions $100\text{mm}\times 100\text{mm} \times 27\text{mm}$ at a constant speed. A motorised traverse with a stepper motor was used to vary the speed of the plate from $100\mu$m/s to $2$cm/s. A DM542 digital microstepper driver was connected to a computer through a data acquisition system from National Instruments. To ensure that a contact line is present in all the experiments, we only allowed the plate to dip into the liquid bath. This restricts our experiments only to advancing cases. To prevent contamination, the glass plate and the tank were thoroughly cleaned with isopropyl alcohol followed by distilled water, and dried before each experiment. The meniscus was illuminated using a thin laser sheet (thickness of approximately 0.5mm) which was created by a combination of biconcave and cylindrical plano-convex lenses from a 532nm 2W diode laser. The flow was seeded with $5\mu$m polyamide particles for all the experiments. The Stokes number for the particles over the entire range of velocities and fluid viscosities ranged from $3.2\times 10^{-7}$ to $6.6\times 10^{-4}$, with particle relaxation times always less than $0.2\mu$s, thus ensuring that the particles faithfully followed the flow. All PIV images were captured using a Photron Fastcam Nova S9 high-speed camera connected to a macro lens. The images were captured at frame rates ranging from 10fps to 1000fps depending on the speed of the plate. To ensure that the surface level of the fluid did not increase when the plate was dipped into the liquid bath, a programmable syringe pump was employed to withdraw fluid at a prescribed flow rate from the bottom of the tank. In all the experiments, the interface shape and the flow settled into a steady state after a short transient. Only the PIV images after this initial transient were processed. The steady nature of the flow allowed us to also generate streakline images, and a sample image obtained from the experiment is shown in figure \ref{fig:streaklines-plot}. The absence of crossings in the streaklines indicates that the flow is indeed steady (also see \cite{supplementary_data} for a sample transient image).

A digital delay generator was used to synchronize the laser and the high-speed camera. The field of view of the high-speed camera with the macro lens varied from 4.2mm $\times$ 4.2mm to nearly 8mm $\times$ 8mm with a spatial resolution in the range of $4\mu$m/pixel to nearly $8\mu$m/pixel. Preprocessing of the images was performed prior to the PIV analysis, including average background subtraction, image equalization, and masking. The particle images were analyzed using a multi-grid, window-deforming PIV algorithm. In addition, since the flow is steady, an ensemble PIV correlation was employed to improve the signal-to-noise ratio in a small interrogation window (typically 8 pixels $\times$ 8 pixels).   

As shown in figure \ref{fig:operating_regime_map}, the primary goal of this paper is to investigate advancing contact angles with dynamic contact angle $\theta_d < 90^{\circ}$. This was most easily achieved with the immersion of a glass plate into a bath of silicone oil at low speeds. The properties of the silicone oils used in the experiments are given in table \ref{tab:fluid_properties}. The wetting properties of the glass substrates were also characterised by measuring the static advancing and receding angles with the help of a contact angle meter (Kruss DSA25S). As clearly evident in table \ref{tab:contact_angle_hysteresis}, the static hysteresis in all the experiments was found to be very small.

\begin{table}
  \begin{center}
\def~{\hphantom{0}}
  \begin{tabular}{cccc}
    &       Density                 & Viscosity             & Surface tension \\ 
    &       $\rho_{\text{oil}}$     & $\mu_{\text{oil}}$    & $\gamma$ \\ 
    &         kg/m$^3$  & $10^{-3}$ Pa.s     & $m$N/m \\[3pt]
Air & 1.207 & 0.0189 & - \\ 
10 cSt Silicone oil & 941 & 8.8 & 19.5 \\
20 cSt Silicone oil & 950 & 17.97 & 19.6 \\
100 cSt Silicone oil & 960 & 90.45 & 19.9\\
500 cSt Silicone oil & 965 & 516.67 & 18.5\\
  \end{tabular}
   \caption{Properties of fluids used in experiments}
   \label{tab:fluid_properties}
  \end{center}
\end{table}

\vspace{15mm}

\begin{table}
 \begin{center}
  \begin{threeparttable}
\begin{tabular}{ p{4.5cm}  c  c  c } 
   & $\theta_{s,a}$ (\text{deg.})   & $\theta_{s,r}$ (\text{deg.})   & Hysteresis* (deg.)\\[3pt]
10 cSt Silicone oil      & 6.8 $\pm$ 2      & 5.3$\pm$1.4     & 1.5\\
100 cSt Silicone oil     & 10.4 $\pm$1.6    & 8.7 $\pm$ 1.1   & 1.7\\
500 cSt Silicone oil     & 11.9 $\pm$ 0.7   & 9.2 $\pm$ 0.7   & 2.7\\
\end{tabular}
    \begin{tablenotes}
      \footnotesize
      \item *Since there is an overlap in the error bars in the receding and advancing angles, only the mean value of the hysteresis is reported.
    \end{tablenotes}
   \end{threeparttable}
 \caption{Characterising the hysteresis for silicone oils on glass substrates}
 \label{tab:contact_angle_hysteresis}
 \end{center}
\end{table}

\begin{figure}
\centering
\includegraphics[trim = 0mm 0mm 0mm 0mm, clip, angle=0,width=0.95\textwidth]{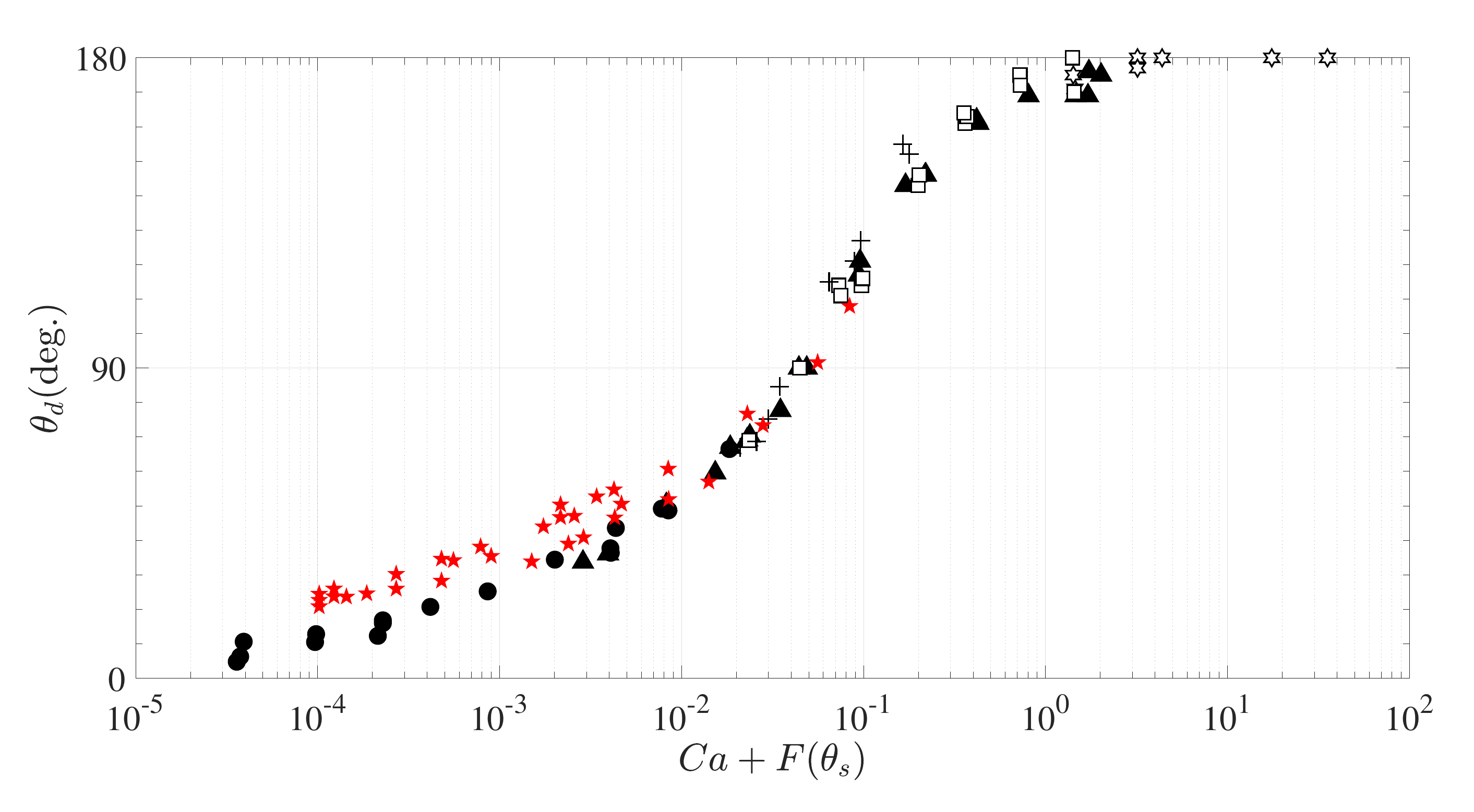}
\caption{Variation of dynamic contact angle, $\theta_d$, with an adjusted capillary number, $Ca + F(\theta_s)$. Here $F(\theta_s)$ represents the shift factor which compensates for the effect of static contact angle in terms of Ca. Shift factor is calculated considering $\theta_d = \theta_s$ and the corresponding value of $Ca$ is assigned to $F(\theta_s)$. All the markers in black are taken from \cite{hoffman1975study} and those in red ({$\color{red}\filledstar$}) represent present data obtained using different grades of silicone oil.}
\label{fig:Hoffmann_plot}
\end{figure}

To determine the interface location and shape, several particle images were combined and the locus of particle streaks at the interface are considered to be representative of the interface itself. The interface angle was determined by fitting a two-term exponential function to the interface data points with an $R$-square value always above $0.99$. The local angle along the interface was then determined from this fit. Dynamic contact angles for various capillary numbers shown in figure \ref{fig:Hoffmann_plot} were compared with the classical study of \cite{hoffman1975study}. Despite the differences in the geometry between the present study and that of Hoffman, the collapse of all contact angle data in the two experiments shows that the relation between dynamic contact angle and capillary number is universal. Fitting the Cox-Voinov model of the form $\theta_d^3-\theta_s^3 = A~Ca$, the data of \cite{hoffman1975study} results in the value of slope $A\approx 81.1$ which is very close to the value 78.7 in the present experiments. More information about comparison with dynamic contact angle models is available in \cite{supplementary_data}.

In the next section, we discuss the theoretical basis for the comparison of flow fields from experiments with models.

\section{\label{sec:theory}Theoretical background}
\begin{figure}
\centering
    \subfigure[]{\label{fig:geometry_huh}
       \includegraphics[trim = 0mm 0mm 0mm 0mm, clip, angle=0,width=0.3\textwidth]{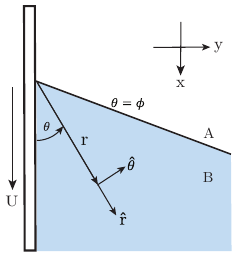} }
    \hspace{5mm}
    \subfigure[]{\label{fig:rolling_flow_schematic}
       \includegraphics[trim = 0mm 0mm 5mm 5mm, clip, angle=0,width=0.2\textwidth]{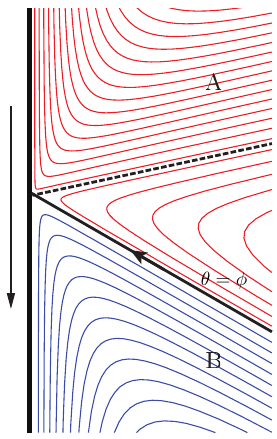} }
    \hspace{5mm}
    \subfigure[]{\label{fig:coordinate_system}
       \includegraphics[trim = 0mm 0mm 0mm 0mm, clip, angle=0,width=0.3\textwidth]{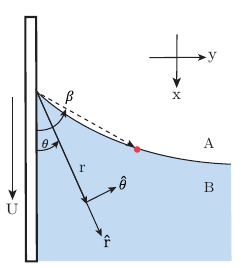} }
\caption{(a) Cylindrical polar coordinate system $(r,\theta)$ used in \cite{huh1971hydrodynamic} for flow in a flat wedge separating two immiscible fluids, A and B, with a constant wedge angle $\phi$, caused by a plate moving at constant speed $U$. (b) Typical flow pattern predicted by Huh \& Scriven's theory when $\lambda \ll 1$. The arrow at the interface represents the direction of motion of fluid particles along the interface. When fluid B undergoes a `rolling' motion, fluid A exhibits a `split-streamline' motion with the splitting streamline shown by a dashed line. (c) Coordinate system for a wedge with a curved interface with angle $\beta$ varying along the interface. }
\label{fig:schematic_analytical}
\end{figure}

The earliest theoretical models for moving contact lines were aimed at developing a local model valid only in the vicinity of the contact line. A schematic of such a `local' framework is shown in figure \ref{fig:geometry_huh} where the dynamics is governed by the Stokes equations for $r \ll L$ with $L$ being the macroscopic length scale in the problem. \cite{huh1971hydrodynamic} were among the first to develop a simple model of a moving contact line assuming the interface to be flat and employing the no-slip condition on the moving wall. Their study showed the presence of a singularity in the shear stress at the contact line. Nevertheless, their model predicts flow fields in both phases that have regular and smooth behaviour away from the contact line. A key prediction in the HS71 model is that the higher viscosity fluid exhibits a rolling motion and the lower viscosity fluid exhibits a split-streamline motion. This means that the fluid particles at the interface move towards the moving contact line in an advancing contact line problem as shown in figure \ref{fig:rolling_flow_schematic}.

To relieve the Huh \& Scriven singularity, slip models (\cite{cox1986dynamics, kirkinis2013hydrodynamic,kirkinis2014moffatt,febres2017existence}) were subsequently developed. \cite{cox1986dynamics}, in a seminal paper, developed an asymptotic model by dividing the flow into three regions: an outer region where the interface shape is affected by the geometry of the problem, a slip-dominated inner region in the vicinity of the moving contact line, and an intermediate region merging these two regions as shown in figure \ref{fig:Cox_matching_regions}. The model of \cite{cox1986dynamics} is valid in the limit of $Re \ll 1$ and $Ca \ll 1$ where $Re = \rho U L/\mu$ and $\mu$ is typically taken to the viscosity of the more viscous phase. This restriction on $Re$ ensures that the flow in the intermediate and inner regions is always in the viscous regime.
By `intermediate' region, we refer to regions of interest that are smaller than the macroscopic length scale, $L$: $l_s \ll l_i < L$. \cite{sibley2015asymptotics} noted that distinct inner and outer regions exist with an intermediate region sandwiched between the two, consistent with figure \ref{fig:Cox_matching_regions}, if the following conditions hold:
\begin{equation}
    r_{\text{outer}} \gg L \mathrm{e}^{-1/|Ca|} \quad \text{\&} \quad r_{\text{inner}} \ll l_s \mathrm{e}^{1/|Ca|} \quad \implies \quad |Ca \ln l_s| \ll 2.
\end{equation}
To access a distinct and wide intermediate region in the experiments, it is therefore sufficient to ensure that the capillary number is always small, typically less than or equal to $10^{-2}$. Since the slip length, $l_s$ is typically in nanometers, the above condition is always satisfied in all our experiments.

\subsection{Models for flow fields} \label{sec:models_flow_fields}
In our experiments, the geometry resembles that of a wedge formed between a vertically moving plate and a curved interface. Our aim is to compare flow fields from experiments in this geometry with well-known models in the literature. Following the seminal work of \cite{moffatt1964viscous} who determined the flow in a wedge formed between two flat plates, \cite{huh1971hydrodynamic} determined the flow in a wedge formed between a moving plate and a flat interface, and was subsequently amended by \cite{chen1997velocity} to incorporate the effects of a curved interface.

\subsubsection{Fixed wedge solution} \label{sec:fixed_wedge}
The `local' solution developed by HS71 assumes the interface to be flat as shown in figure \ref{fig:geometry_huh}. The flow in the vicinity of the moving contact line is governed by the biharmonic equation for the streamfunction \cite{moffatt1964viscous},
\begin{equation}
    \bnabla^4 \psi = 0,
\end{equation}
where 
\begin{equation}\label{eq:Scriven1}
   \psi(r,\theta) = r(a \sin \theta + b \cos \theta + c \theta \sin \theta + d \theta \cos \theta).
\end{equation}

Using conditions of no-slip, no-penetration on the moving solid surface, continuity of tangential velocity, and tangential stress at the interface (no Marangoni effects), the streamfunction reduces to
\begin{equation}\label{eq:streamfunction_HS71}
   \psi(r,\theta;\phi) = rf(\theta,\phi) = r\left(\frac{\phi\sin\theta - \theta\sin\phi \cos(\theta -\phi) }{\phi - \sin\phi \cos\phi}\right).
\end{equation}
In the above expression, the gas above the liquid is assumed to be passive, i.e. $\lambda = 0$. It can be easily shown that the shear stress obtained from the above expression diverges like $1/r$ as $r\rightarrow 0$. In the HS71 framework, the interfacial velocity, $v_i^{HS}$, is identically equal to the radial velocity and is independent of the radial location along the interface and is given by
\begin{equation}
    v_i^{HS} = v_r(r,\phi) = U \left(\frac{\phi \cos \phi - \sin \phi}{\phi - \sin \phi \cos \phi}\right).
    \label{eq:v_int_HS71}
\end{equation}
In experiments, the interface is always curved due to the presence of external forces such as gravity which induces a mean curvature to the interface. To facilitate a comparison of the streamfunction with experiments, we modify \eqref{eq:streamfunction_HS71} for a curved interface. The recipe for doing this was first given by \cite{chen1997velocity}.

\subsubsection{Modulated wedge solution}\label{sec:MWS_theory}

If the interface is curved, shown schematically in figure \ref{fig:coordinate_system}, with the interface angle at any radial position given by $\beta(r)$, then the streamfunction in equn. \eqref{eq:streamfunction_HS71} can be modified by replacing the constant angle $\phi$ with a variable angle $\beta(r)$:
\begin{equation}\label{eq:Modulate_wedge_1}
   \psi(r,\theta;\beta(r)) = rf(\theta,\beta) = r\left(\frac{\beta\sin\theta - \theta\sin\beta \cos(\theta -\beta) }{\beta - \sin\beta \cos\beta}\right).
\end{equation}
This solution, also used by \cite{chen1997velocity} is referred to as the modulated-wedge solution (MWS hereafter). The above solution is identical to the leading order solution in the intermediate region given by \cite{cox1986dynamics}.
As one approaches the contact line, the modulated wedge angle $\beta$ approaches the fixed wedge angle $\phi$. Therefore, as $r \rightarrow 0$, the expressions $\eqref{eq:streamfunction_HS71}$ and $\eqref{eq:Modulate_wedge_1}$ become identical. This allows us to make a three-way comparison between the HS71 solution, modulated wedge solution, and experiments.

The function $\beta(r)$ can be obtained from experiments and inserted into \eqref{eq:Modulate_wedge_1} to obtain streamfunction everywhere in the fluid domain. 
The radial and tangential velocities can also be easily computed as
\begin{eqnarray}
   && v_r(r,\theta;\beta) = \frac{\partial f}{\partial\theta}, \\
   && v_{\theta}(r,\theta;\beta) = -f - r\frac{\partial f}{\partial \beta}\frac{\partial \beta}{\partial r},
\end{eqnarray}
Using the expression for $f(\theta,\beta)$ from \eqref{eq:Modulate_wedge_1}, we have
\begin{align} 
\frac{v_r(r,\theta;\beta)}{U} = &\frac{\beta \cos\theta + \theta \sin \beta \sin (\theta - \beta) - \sin \beta \cos (\theta-\beta)}{\beta - \sin\beta \cos\beta}, \label{eq:vr_MW}\\ 
\frac{v_{\theta}(r,\theta;\beta)}{U} = &\frac{-\beta \sin\theta + \theta\sin \beta \cos (\theta-\beta)}{\beta - \sin\beta \cos\beta} \nonumber
   \\ &  + r\frac{d\beta}{dr} \left(\frac{2\sin^2\beta(\beta \sin\theta - \theta\sin\beta \cos(\theta-\beta))}{(\beta - \sin\beta \cos\beta)^2} - \frac{\sin\theta -\theta \cos(2\beta - \theta)}{\beta - \sin\beta \cos\beta}\right). \label{eq:vtheta_MW}
\end{align}
Unlike the HS71 solution, the interfacial velocity now depends on both the radial and the angular velocities. Setting $\theta = \beta$ in equns. \eqref{eq:vr_MW} and \eqref{eq:vtheta_MW}, we get
\begin{equation}\label{eq:modulated_interfacial_speed}
    v_i^{MWS} = v_r(r,\beta) \cos(\alpha-\beta) + v_{\theta}(r,\beta) \sin(\alpha-\beta),
\end{equation}
where $\alpha(r)$ is a measure of the slope of the interface and is related to $\beta(r)$ by the expression
\begin{equation}
    \alpha(r) = \beta(r) + \tan^{-1}\left(r\spopa{\beta}{r}\right).
\end{equation}
After simplification, the expression \eqref{eq:modulated_interfacial_speed} reduces to
\begin{equation} \label{eq:v_int_MWS}
    v_i^{MWS} = \frac{U}{\cos(\alpha - \beta)}\left(\frac{\beta \cos \beta - \sin \beta}{\beta - \sin \beta \cos \beta}\right).
\end{equation}
At the contact line, $\alpha=\beta=\phi$. Equation \eqref{eq:v_int_MWS} reduces to HS71 solution in \eqref{eq:v_int_HS71} as $r\rightarrow 0$. But for other radial positions, the interfacial speed $v_i^{MWS}$ will no longer be the same as $v_i^{HS}$ but will increase in magnitude while moving away from the contact line and then become constant as the change in the difference ($\alpha-\beta$) becomes small.

The expression for interface shape, expressed in terms of $\beta(r)$, can fully describe the flow field using the streamfunction \ref{eq:Modulate_wedge_1}. In the next section, we apply analytical models and show how interface shapes in the intermediate and outer regions can be compared against experiments.

\subsection{Models for complete interface shape}\label{sec:Interface_Shape_theory}
As discussed in the previous sections, if the interface shape, expressed either in terms of the local slope, $\alpha(r)$, or in terms of a variable angle $\beta(r)$, is available, then the full flow field can be determined using the modulated wedge solution. We employ two distinct theoretical models to obtain the shape of the interface shape in the intermediate and outer regions. The first approach relies on developing a composite solution for the interface shape by using the Cox model. This composite solution is parameterised by a single scalar parameter which is determined by matching the solution to experiments. The second approach uses a generalisation of lubrication equations and reduces the problem to a system of coupled differential equations.

Recall that in the three-region framework of \cite{cox1986dynamics}, the inner slip-dominated region matched to an outer region via an intermediate region. Since the inner region is often beyond the resolution of any experiment, a direct comparison of interface shape with the Cox model with experiments is likely to result in poor agreement. Viscous bending, a phenomenon where the interface deforms sharply in the vicinity of the inner region, is a direct function of the capillary number. At high capillary numbers, i.e. when $Ca \sim O(1)$, viscous bending is likely to persist over a large length scale, typically of $O(300-500)\mu$m. But the challenge is particularly acute when $Ca \ll 1$ where the length scale of viscous bending can be $O(10)\mu$m or smaller. To improve the prediction of interface shape over all length scales, it is necessary to incorporate the effect of the outer region which may be dominated by other forces such as gravity. \cite{dussan1991} precisely carried out such an extension, referred to as the DRG model hereafter.

\begin{figure}
\centering
\subfigure[]{
\label{fig:f0men}
\includegraphics[trim = 0mm 0mm 0mm 0mm, clip, angle=0,width=0.4\textwidth]{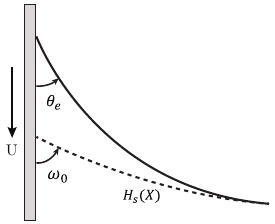}
}
\subfigure[]{
\label{fig:GLM_Schematic}
\includegraphics[trim = 0mm 0mm 0mm 0mm, clip, angle=0,width=0.45\textwidth]{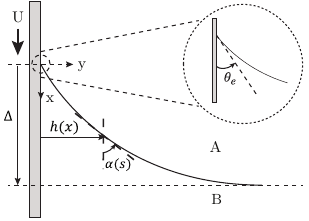}}
\caption{(a) The geometrical set-up for the DRG model with the outer solution, $H_s(X)$ making an angle $\omega_0$ at the moving wall and $\theta_e$ is the microscopic contact angle in the inner region as per \cite{cox1986dynamics}. (b) The dynamic meniscus in the GLM framework with $\alpha(s)$ being the local angle along the curved interface, $s$ is the arc length measured from the contact line, $h(x)$ is the position of the interface measured from the moving plate and $\Delta$ is the position of the contact line with respect to the horizontal level of the interface in the far field.
}
\end{figure}

In the DRG model, the problem of finding the full interface shape is reduced to a search for a single algebraic parameter, termed $\omega_0$. In the absence of motion, an interface deforms near the plate forming a static meniscus of characteristic length $l_c$, the capillary length. According to the classical three-layer model of \cite{cox1986dynamics}, the interface shape in the intermediate region is given by
\begin{equation}\label{eq:intermediate}
 g(\alpha)\sim d_0 + Ca\ln(r/l_c), 
\end{equation}
where the function $g(x)$ is given by
\begin{eqnarray}\label{eq:gtheta}
  g(x) = \int^x_0 \frac{t - \cos t \sin t}{2\sin t} dt,
\end{eqnarray}
and
\begin{equation}\label{eq:inner_outer}
 d_0 = g(\omega_0)=g(\theta_{e}) + Ca\ln(l_c/l_s).
\end{equation}
The viscosity ratio, $\lambda$, has been assumed to be negligible in the above expression for simplicity, though it can be included as per the full Cox model without any difficulty. In \eqref{eq:inner_outer}, $l_s$ is the slip length, $\theta_{e}$ is the microscopic contact angle, and $\omega_0$ plays the same role as the apparent contact angle. Following \cite{cox1986dynamics}, matching the solutions between the intermediate region and the inner region (see figure \ref{fig:Cox_matching_regions}), we obtain
\begin{equation}\label{eq:intermediate_inner}
 g(\alpha) = g(\theta_{e}) + Ca\ln(r/l_s).
\end{equation}
\cite{dussan1991} showed that $\omega_0$ can be determined by matching the interface shape from the analytical model with experimental data. In the outer region, the interface shape is dictated by the shape of the static interface such that it matches the solution in the intermediate region. For example, the shape of a static meniscus would be determined not only by the far-field boundary conditions where the slope of the interface vanishes but also by a boundary condition on the plate. In the same way, the outer solution will have to be parameterised by an `effective' contact angle boundary condition which is equal to $\omega_0$. This is shown schematically in figure \ref{fig:f0men}. Let us assume that the interface shape in the outer `static' region assumes the form
 \begin{equation}\label{eq:outer}
    \theta_{s} = f_0\left(\frac{r}{l_c};\omega_0\right),
 \end{equation}
where $\theta_{s}$ is a local slope estimated along the static interface. The exact form of $f_0$ for a 2D static meniscus is given in section \ref{sec:interface_shape}. Matching the intermediate solution \eqref{eq:intermediate} to the outer solution above, we obtain an expression for the interface angle given by
\begin{equation}\label{eq:Cox_intermediate}
  g(\alpha)=g(\omega_0) + Ca\ln\left(\frac{r}{l_c}\right).
\end{equation}
The composite DRG solution is obtained by adding the static and Cox solutions, i.e, \eqref{eq:outer} and \eqref{eq:Cox_intermediate}, to give
 \begin{equation}\label{eq:drgcomp}
   \alpha(r) = g^{-1}\left[g(\omega_0)+Ca\ln\left(\frac{r}{l_c}\right)\right] + \left(\theta_s - \omega_0\right).
 \end{equation}
The parameter, $\omega_0$, can be interpreted as an empirical parameter that can be evaluated by matching \eqref{eq:drgcomp} with experimentally obtained interface shape. 

%

An alternative approach to obtain the full interface shape was proposed by Snoeijer and coworkers. This is a simple and direct approach involving the generalized lubrication equations used in free-surface flows even for problems with large interface slopes \citep{snoeijer2006free, chan2013hydrodynamics, chan2020cox}. This is achieved by expanding the Stokes equations about flow in a wedge with a constant wedge angle \citep{huh1971hydrodynamic}, perturbed around $Ca\ll 1$, i.e. variations of the interface slope are assumed to be slow. The model is described in terms of two coupled differential equations for the interface shape, $h(s)$, and local interface angle, $\alpha(s)$, as shown schematically in figure \ref{fig:GLM_Schematic}.
\begin{subequations}
  \begin{align}
    \spopasq{\alpha}{s} = &~\frac{3 Ca}{h(h+cl_s)} f(\alpha,\lambda) - \frac{1}{l_c^2}\cos(\alpha),\\
    \spopa{h}{s} = &~\sin(\alpha),
 \end{align}
 \label{eq:glm}
\end{subequations}
where $s$ is the arc length along the interface and $c$ is a constant chosen to match with the slip region. For small contact angles ($\theta_e\ll 1$) and free-surface flows ($\lambda=0$), one can take $c=3$ \citep{chan2020cox}. In the limit of $\lambda=0$, the function $f(\alpha,\lambda)$ is given by:
\begin{equation}
    f(\alpha,0)=-\frac{2\sin^3(\alpha)}{3(\alpha-\sin(\alpha)\cos(\alpha))}.
\end{equation}
Integrating \eqref{eq:glm} and using $Ca\ll 1$, \cite{snoeijer2006free} showed that the structure of \eqref{eq:intermediate_inner} can be recovered. In contrast to the method of matched asymptotics involving matching solutions from distinct regions \citep{cox1986dynamics,dussan1991}, the generalised lubrication model (hereafter referred to as GLM) in equations \eqref{eq:glm} gives a full description of the interfacial profile in a more convenient way. For the present experiments, the GLM equations are solved using the following boundary conditions on the moving plate and the far-field:
\begin{eqnarray}
    \alpha(s=0)=\theta_e;\quad \alpha(s\to\infty)=\frac{\pi}{2}, \qquad h(s=0)=0. \label{eq:glmbc}
\end{eqnarray}
Detailed comparisons of the interface shape from the above two models with experiments are discussed in section \ref{sec:interface_shape}.


\section{\label{sec:results}Results}

The results are broadly divided into three parts. First, in \S\ref{sec:interface_shape}, we extract the interface shape from experiments and compare it against theoretical models. Second, in \S\ref{sec:compare_modulated_wedge_solution}, we extract flow fields from the experiments and compare them against the modulated wedge flow solutions described in \S\ref{sec:MWS_theory}. Finally, the velocity at the interface is compared against theoretical models in \S\ref{sec:interface_speed}.

\subsection{Interface shape} \label{sec:interface_shape}
Interface shape has a direct bearing on the flow on either side of it, and it is therefore necessary to investigate models for interface shapes carefully as is done below. 

\begin{figure} 
\centering
\includegraphics[trim = 0mm 0mm 0mm 0mm, clip, angle=0,width=0.65\textwidth]{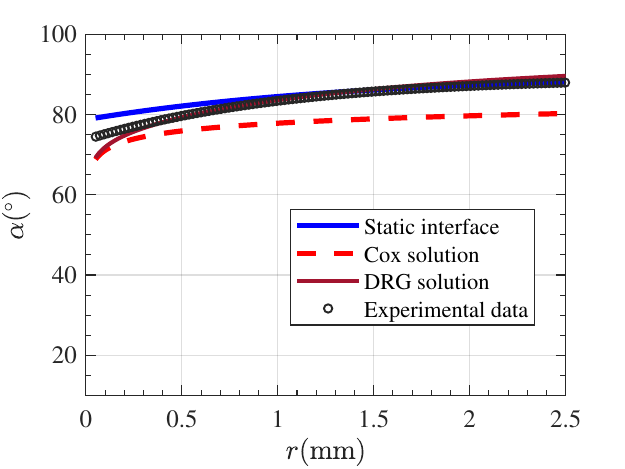}
\caption{Comparison of interface shape obtained in the experiments against the DRG solution, Cox's intermediate solution, and the static (outer) solution. This solution corresponds to the interface between air and 500 cSt silicone oil with $Re=2.61 \times 10^{-3}$ and $Ca = 2.79\times 10^{-2}$. $\omega_0$ is obtained by fitting the DRG model to experiments and is approximately $78.8^{\circ}$.} 
\label{fig:drg_comparison}
\end{figure}

For the DRG model (refer to \S \ref{sec:Interface_Shape_theory} for more details), we require knowledge of interface shape in the outer static region. For the present problem of a flat plate vertically advancing into a liquid bath, the interface shape can be written analytically in terms of the local interface angle, $\theta_s$, where the subscript $s$ refers to the static solution. The static shape, written in parametric form is given by
\begin{equation}
    \theta_s = \frac{\pi}{2} - \tan^{-1}\left(\frac{dH_s}{dX}\right)
  \label{eq:f0}
\end{equation}
where $X$ and $H_s$ are the non-dimensional vertical and horizontal coordinates of the static interface with the origin at the contact line such that $X \in [0,X_0]$. An analytical solution for the full nonlinear Young-Laplace equation is readily available and can be written as
\begin{equation}
  H_s(X) = \cosh^{-1}\left(\frac{2}{X_0-X}\right) - \cosh^{-1}\left(\frac{2}{X_0}\right) - \left(4 - (X_0-X)^2\right)^{1/2} + \left(4 - X_0^2\right)^{1/2},
  \label{eq:static_shape_parametric}
\end{equation}
where
\begin{equation}
  X_0 = \sqrt{2}\left(1-\sin\omega_0\right)^{1/2}. 
\end{equation}
Instead of prescribing the static contact angle at $x=0$, we set the angle to be $\omega_0$, an empirical parameter whose value is determined by iteratively minimizing the root mean square error between the interface shape predicted by the DRG model, \eqref{eq:drgcomp}, and the experimental interface shape. The outcome of this procedure is shown in figure \ref{fig:drg_comparison} where the static (outer) solution \eqref{eq:f0}, the Cox (intermediate) solution \eqref{eq:Cox_intermediate}, and the DRG solution are compared against experimental data. The Cox solution shows a prominent viscous bending near the contact line which was not found in the experiments. However, the DRG solution follows the experimental data away from the contact line all the way into the static region. The region where the DRG or experiment data deviates from the static region is a measure of viscous deformation near the wall. Note that viscous deformation is distinct from viscous bending. In the case shown in figure \ref{fig:drg_comparison}, the viscous deformation is close to 1000 $\mu$m, which is clearly comparable to the length scale of the outer solution. Such large viscous deformations are consistent with similar values reported by \cite{rame1996microscopic}.

\begin{figure}
\centering
\includegraphics[trim = 0mm 0mm 0mm 0mm, clip, angle=0,width=0.65\textwidth]{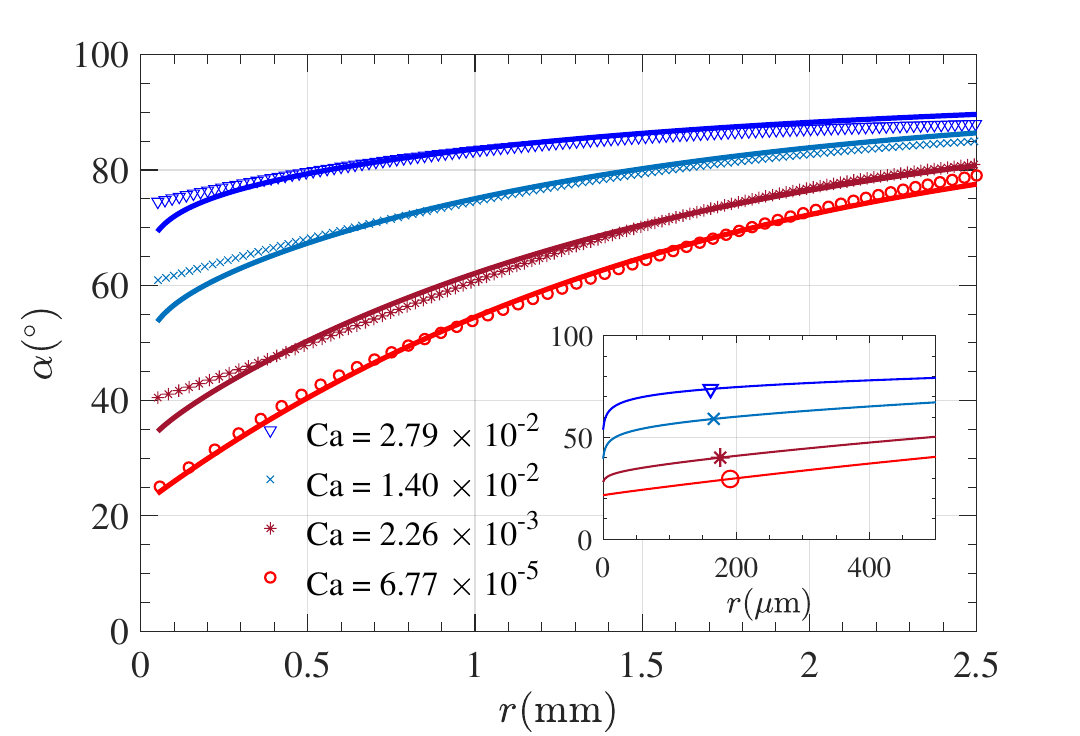}
\caption{Comparison of interface shapes from the experiments with DRG model for four different cases.~~~~ 
$\color{blue}\triangledown$: 500 cSt silicone oil with $Ca=2.79\times10^{-2}$, $\omega_0 \approx 78.8^{\circ}$;~~~~ $\color{d_cyan}\boldsymbol{\bigtimes}$: 500 cSt silicon oil with $Ca=1.4\times10^{-2}$, $\omega_0 \approx 60.9^{\circ}$;~~~~ $\color{maroon}\mathlarger{\mathlarger{\mathlarger{\ast}}}$: 100 cSt silicone oil with $Ca=2.26\times10^{-3}$, $\omega_0 \approx 36.5^{\circ}$;~~~~ 
$\color{red}\boldsymbol{\medcirc}$: 10 cSt silicone oil with $Ca=6.77\times10^{-5}$, $\omega_0 \approx 22^{\circ}$. The inset shows the extent of \emph{viscous bending} described by the DRG model near the contact line.}
\label{fig:DRG_interface_shape}
\end{figure}



\begin{figure}
\centering
\includegraphics[trim = 0mm 0mm 0mm 0mm, clip, angle=0,width=0.65\textwidth]{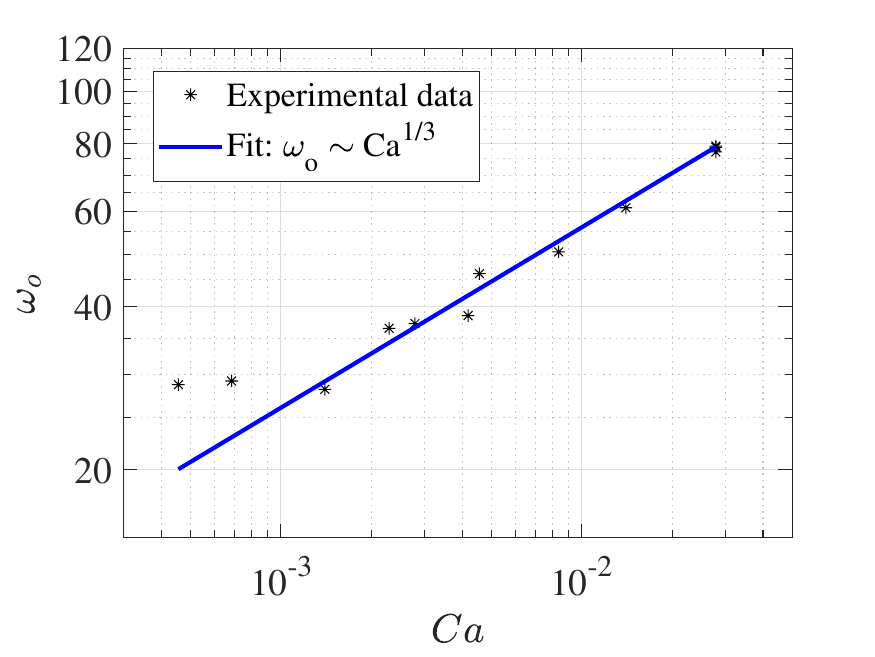}
\caption{Variation of $\omega_0$ with capillary number $Ca$ obtained from DRG model from experiments involving fluids with viscosity ranging from 100 cSt to 500 cSt.}
\label{fig:omega_Ca_figure}
\end{figure}

Figure \ref{fig:DRG_interface_shape} shows a comparison of the DRG model with the experimental interface shapes for 500 cSt, 100 cSt and 10 cSt silicone oils. At low $Ca$ as is the case with 10 cSt oil, the agreement between the DRG model and experiment is very good. The agreement becomes less favourable at higher $Ca$ due to the pronounced viscous bending predicted by Cox's viscous theory. The extent of viscous bending predicted by Cox's theory appears to range from a few microns at low $Ca$ and increases to about 400 $\mu$m for 500 cSt oil. The variation of fitting parameter $\omega_0$ in each of the cases shown in figure \ref{fig:DRG_interface_shape} can also be predicted using the model equation \eqref{eq:inner_outer}. For small to moderate angles, the function $g(\theta) \approx \theta^3/9$. This simplifies equn. \eqref{eq:inner_outer} to the form
\begin{equation}
  \omega_0^{3} \approx  \theta_e^3 + 9 Ca\ln\left(l_c/l_s\right).
\label{eq:Omega_equation}
\end{equation}
The equilibrium angle $\theta_e$ is the static advancing angle given in Table \ref{tab:contact_angle_hysteresis}.
For small equilibrium angles, i.e., $\theta_e \ll 1$, the above expression further simplifies to
\begin{eqnarray}
  \omega_0^3 \sim 9Ca\ln\left(l_c/l_s\right).
  \label{eq:omega_0_vs_Ca}
\end{eqnarray}
If $l_c$ and $l_s$ are nearly constant, as is the case in the present study for Silicone oils of different viscosities, \eqref{eq:omega_0_vs_Ca} predicts that $\omega_0$ has a $Ca^{1/3}$ dependence on the capillary number. Figure \ref{fig:omega_Ca_figure} displays the relationship between $\omega_0$ and $Ca$, consistent with the scaling in equation \eqref{eq:omega_0_vs_Ca}.  

\begin{figure}
\centering
\includegraphics[trim = 0mm 0mm 0mm 0mm, clip, angle=0,width=0.7\textwidth]{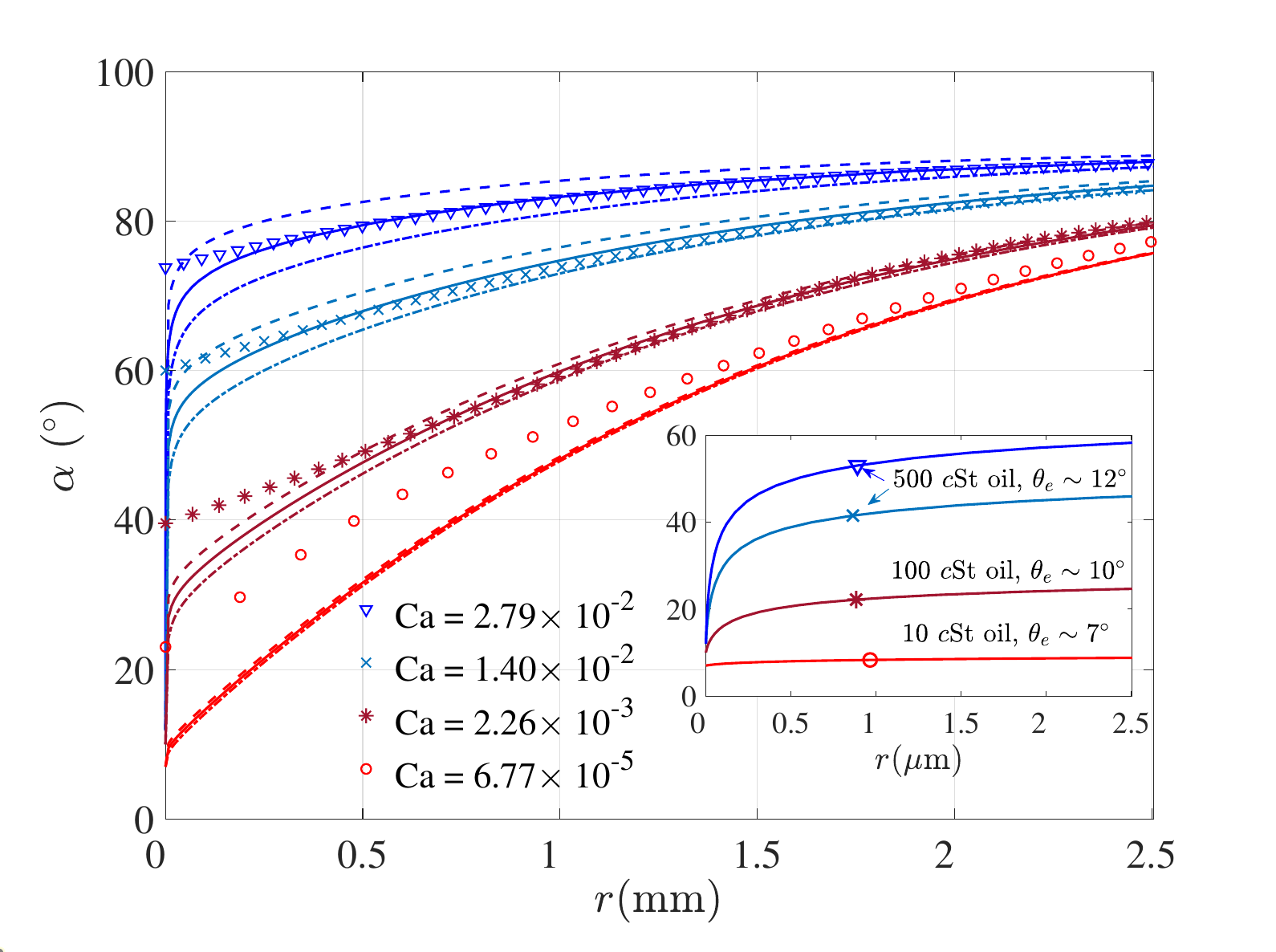}
\caption{Comparison of experimental interfacial shapes (symbols) with GLM predictions (curves) using three different values for the slip length:~~~~ 
$\color{blue}\triangledown$: 500 cSt silicone oil with $Ca=2.79\times10^{-2}$;~~~~ 
$\color{d_cyan}\boldsymbol{\bigtimes}$: 500 cSt silicon oil with $Ca=1.4\times10^{-2}$;~~~~ 
$\color{maroon}\mathlarger{\mathlarger{\mathlarger{\ast}}}$: 100 cSt silicone oil with $Ca=2.26\times10^{-3}$;~~~~ 
$\color{red}\boldsymbol{\medcirc}$: 10 cSt silicone oil with $Ca=6.77\times10^{-5}$. 
GLM predictions with $l_s=2.8$ nm (dashed), $l_s=14$ nm (solid), $l_s=56$ nm (dash-dot). The inset shows \emph{viscous bending} predicted by GLM near the contact line with $l_s=14$ nm.}
\label{fig:Chan_expt_comparison}
\end{figure}

A more direct approach to obtain the interface shapes is to use the recently developed GLM model given by equations \eqref{eq:glm}. The only unknown parameter in this model is the slip length, $l_s$, which can be determined by matching the interface shape from the model against experimental data. The equilibrium contact angle, $\theta_e$ can be obtained from direct measurements given in Table \ref{tab:contact_angle_hysteresis}. Figure \ref{fig:Chan_expt_comparison} shows a comparison of interface shapes between experiments and GLM predictions for three different slip lengths, 2.8 nm, 14 nm and 56 nm. In most cases, a slip length of 14 nm fits the experimental data with reasonable accuracy. A constant slip length, which does not vary with capillary number suggests that the slip length is a fundamental property of the fluid-solid combination. This is consistent with several other studies that measured slip length and found the values in the nanometer range and also independent of the applied strain-rate as discussed in the work of \cite{joseph2005direct}.

The GLM model performed well for most cases except for very low capillary number experiments. At very low $Ca$, viscous effects are confined to a very small region near the wall and the interface shape is dominated by the static shape. This is also consistent with the DRG model shown in figure \ref{fig:DRG_interface_shape} where the static shape largely governs the whole interface shape when $Ca$ is very small. But in the GLM model, viscous effects are assumed to be valid over a much larger length scale and thus could be the reason why there is a deviation at low $Ca$.


\subsection{A comparison of streamfunction with modulated wedge solution}
\label{sec:compare_modulated_wedge_solution}

A key result of this paper is to obtain flow fields in the vicinity of a moving contact line and compare them against existing theories. As discussed in \S\ref{sec:models_flow_fields}, a direct comparison of experimental flow field data cannot be performed with the fixed wedge theory of \cite{huh1971hydrodynamic}, and hence the modulated wedge solution proposed by \cite{chen1997velocity} is adopted in the present study. The effect of the viscosity ratio was neglected in formulating the modulated wedge solution since all experiments were conducted only in the limit of $\lambda \ll 1$. We have checked to ensure that ignoring the viscosity ratio in the theory has minimal effect on the values of the streamfunction. However, the effect of the viscosity ratio and speed of the plate directly affects the interface shape, $\beta(r)$, which in turn modifies the flow fields.

The interface shape from experiments, either in terms of $\alpha(r)$ or $\beta(r)$, is directly used to determine the streamfunction in the bulk using equn. \eqref{eq:Modulate_wedge_1} as discussed in \S\ref{sec:MWS_theory}. Using the quantitative data of velocity field from PIV experiments, streamfuntion fields were determined by using mass conservation. To facilitate direct comparison with theory, the streamfunction value at the moving wall was set to zero. This allows a direct test of viscous theories by not only checking the qualitative shape of the streamlines but also a quantitative comparison of the streamfunction values. 
\begin{figure}
\centering
\subfigure[]{
\includegraphics[trim = 0mm 0mm 0mm 0mm, clip, angle=0,width=0.45\textwidth]{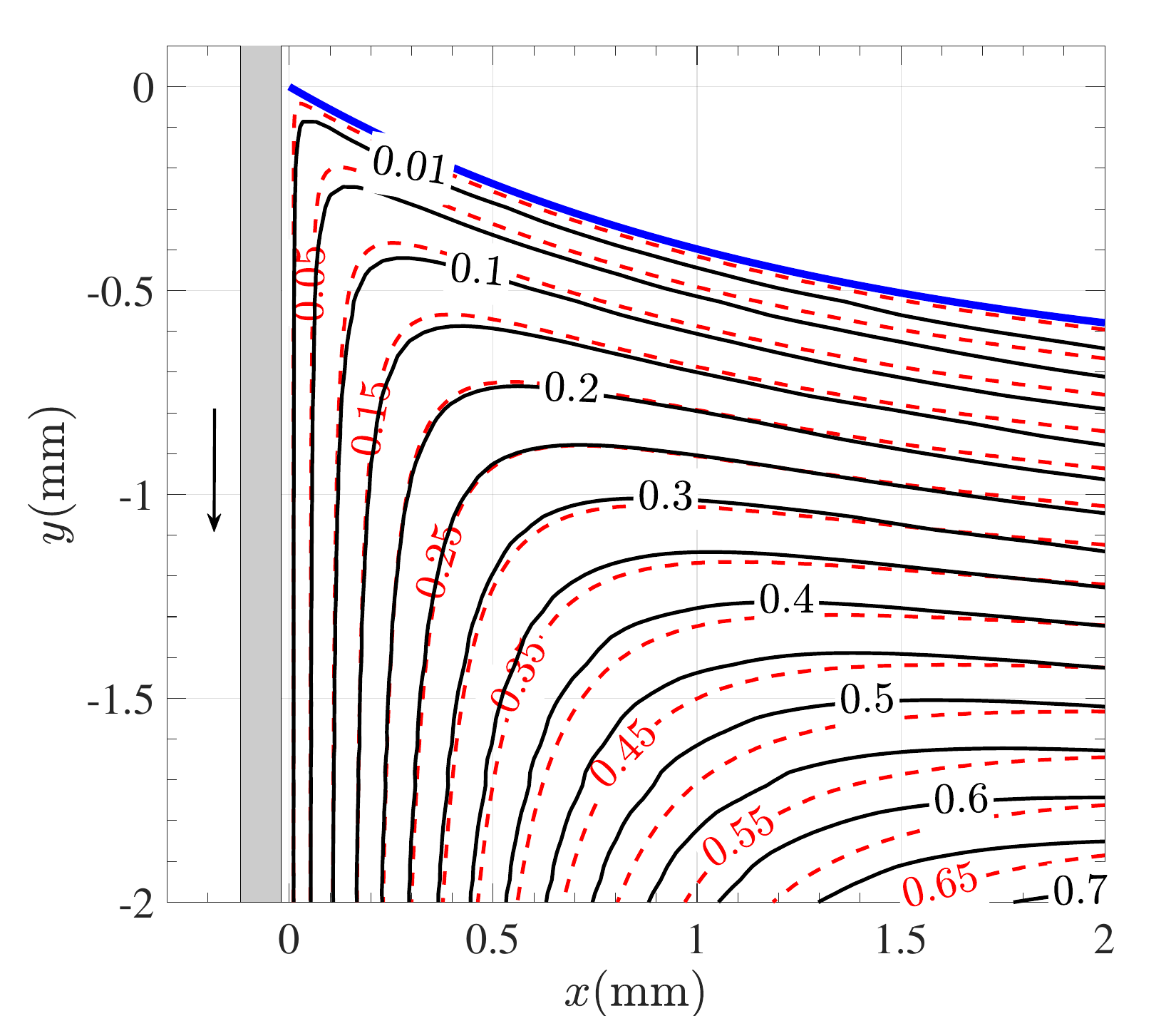}
\label{fig:streamfunction_500cSt}
}
\subfigure[]{
\includegraphics[trim = 0mm 0mm 0mm 0mm, clip, angle=0,width=0.45\textwidth]{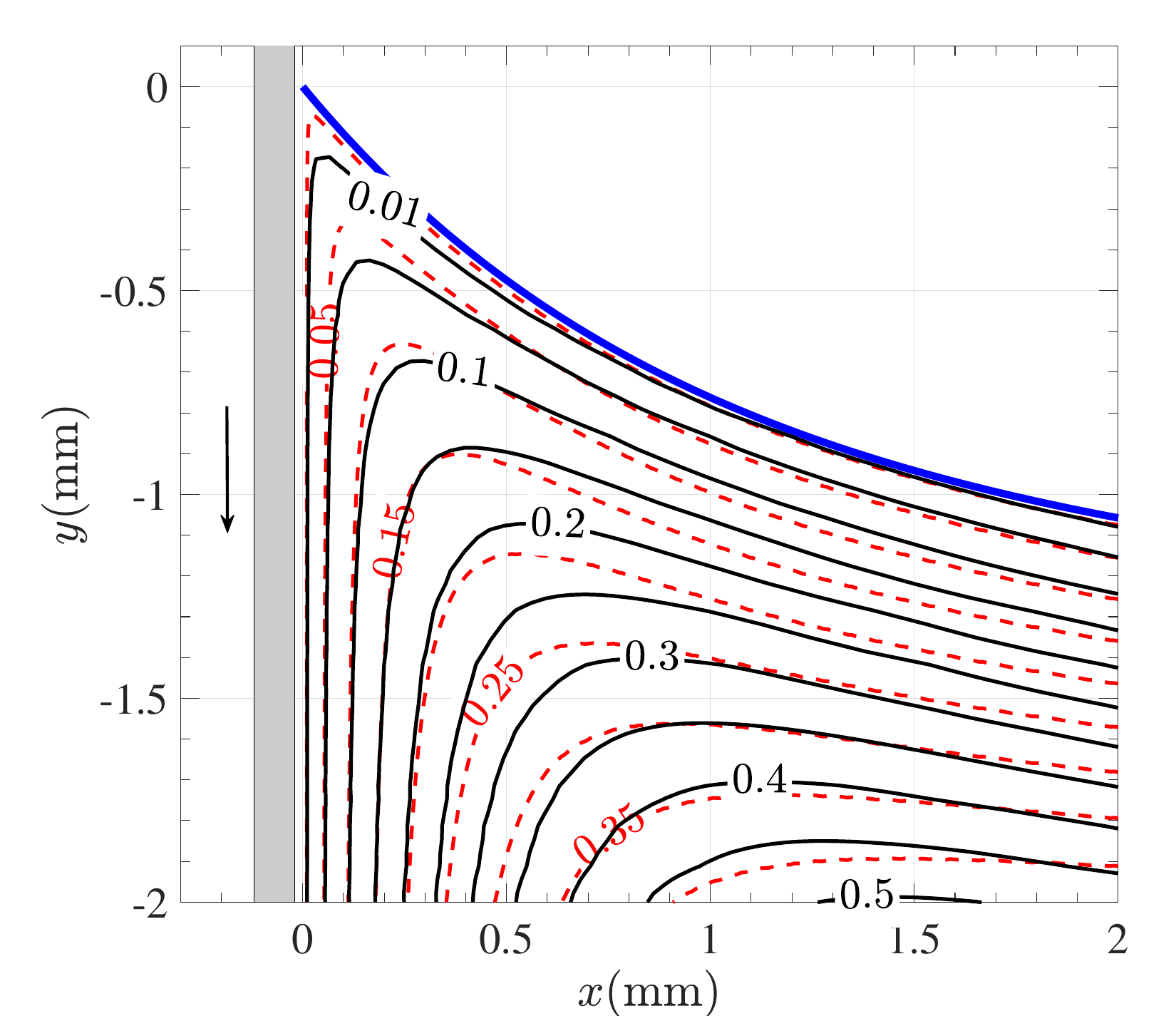}
\label{fig:streamfunction_100cSt}
}
\subfigure[]{
\includegraphics[trim = 0mm 0mm 0mm 0mm, clip, angle=0,width=0.45\textwidth]{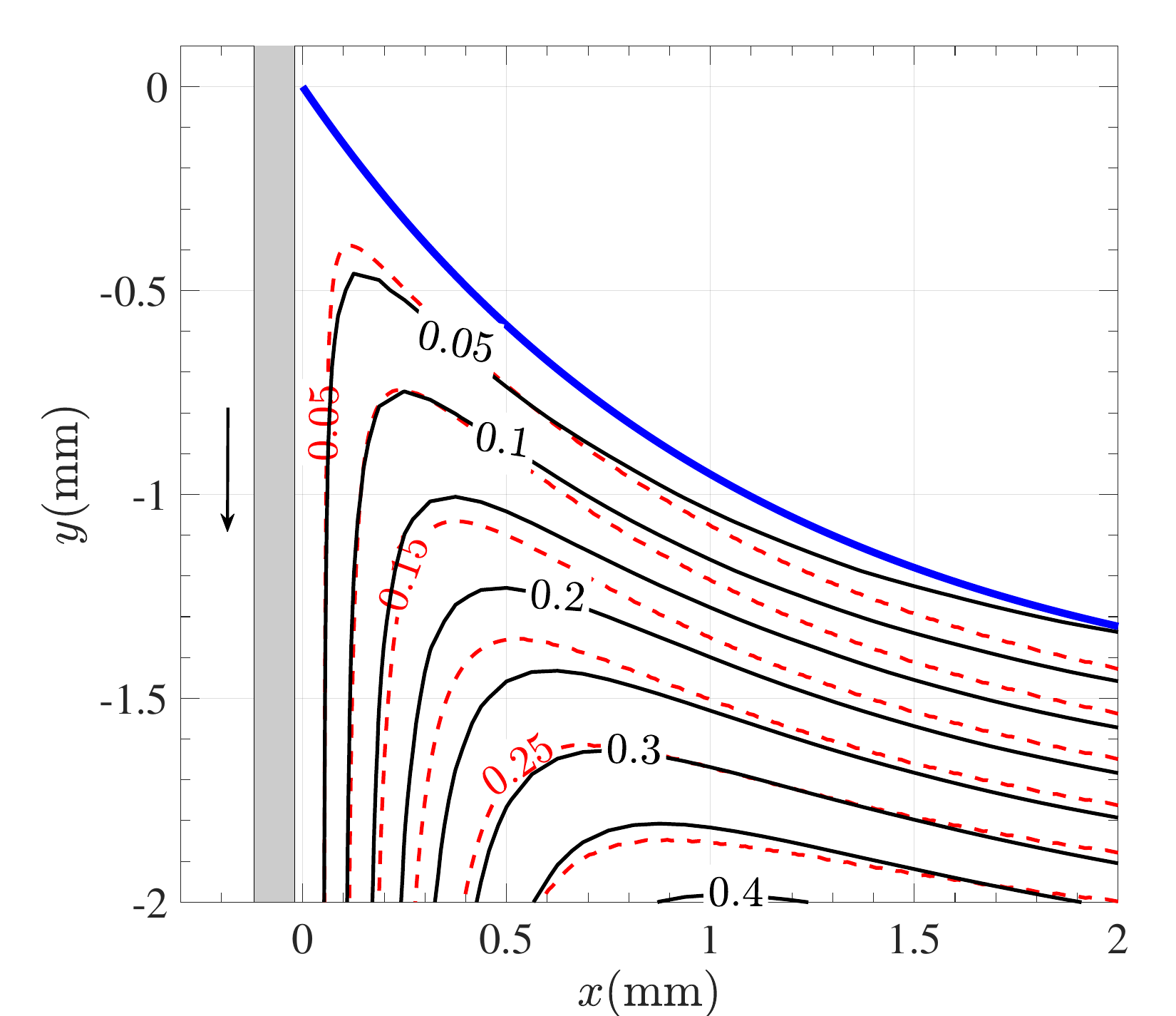}
\label{fig:streamfunction_20cSt}
}
\subfigure[]{
    \includegraphics[trim = 0mm 0mm 0mm 0mm, clip, angle=0,width=0.45\textwidth]{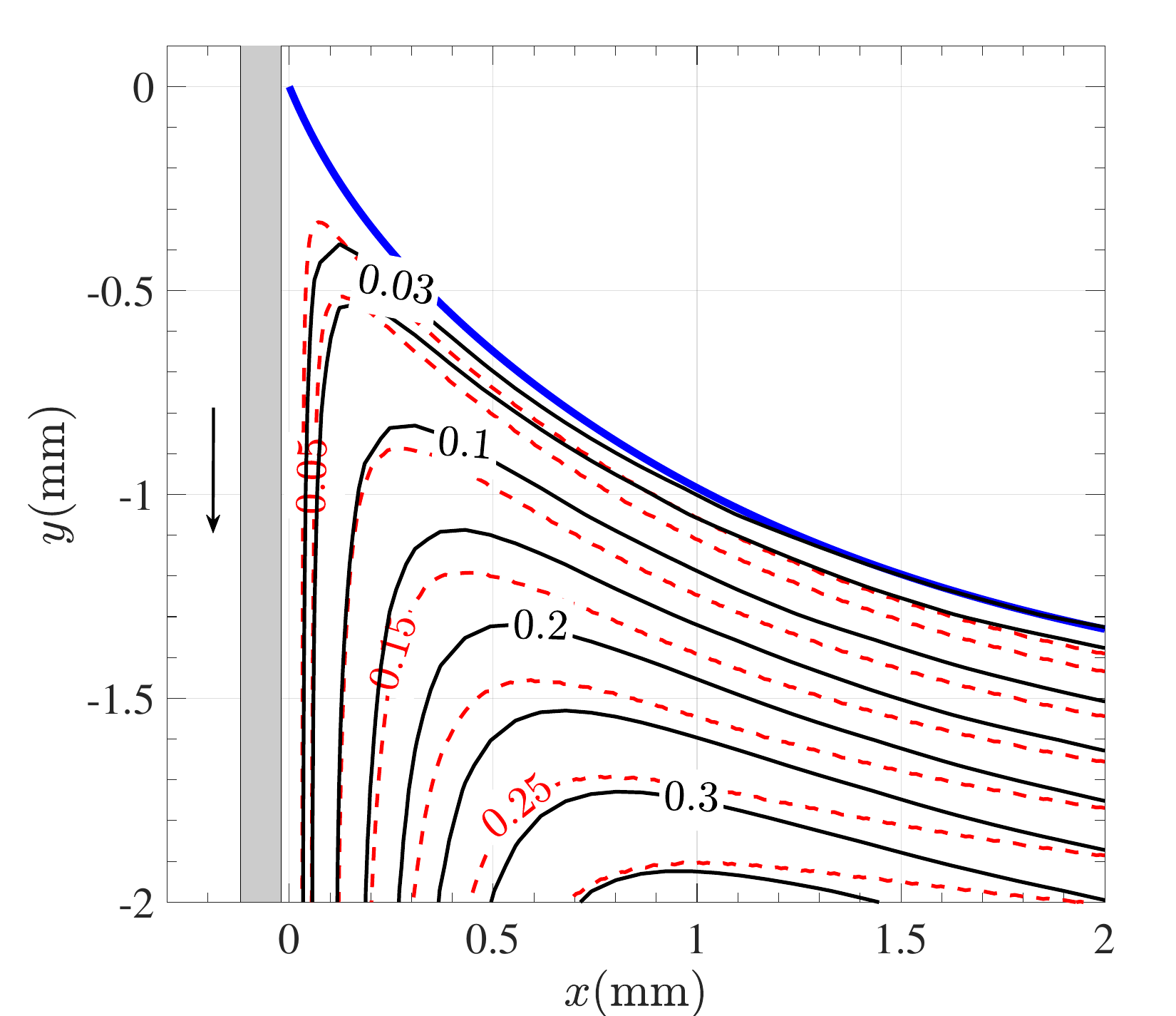}
    \label{fig:streamfunction_10cSt}
}
\caption{Contours of streamfunction obtained from experiments and viscous theory. The gray rectangle represents the solid plate moving downwards, and the blue solid curve represents the interface between air and silicone oil. Experimental streamfunction contours (black solid curves) are derived using mass balance considerations from velocity vector field using PIV experiments and the theoretical predictions (red dashed curves) are obtained from viscous theory incorporating the effect of a curved interface.~~~
(a) Silicone 500 cSt at $Re = 1.3\times 10^{-3}$ and $Ca = 1.4\times 10^{-2}$,~~ (b) Silicone 100 cSt at $Re = 7.7\times 10^{-3}$ and $Ca = 2.41\times 10^{-3}$,~~ (c) Silicone 20 cSt at $Re = 1.15\times 10^{-2}$ and $Ca = 1.36\times 10^{-4}$,~~ (d) Silicone 10 cSt at $Re = 0.023$ and $Ca = 6.77\times 10^{-5}$.}
\label{fig:silicone_streamfunction_varying_cSt}
\end{figure}


\begin{figure}
\centering
\includegraphics[trim = 4mm 1mm 15mm 5mm, clip, angle=0,width=0.65\textwidth]{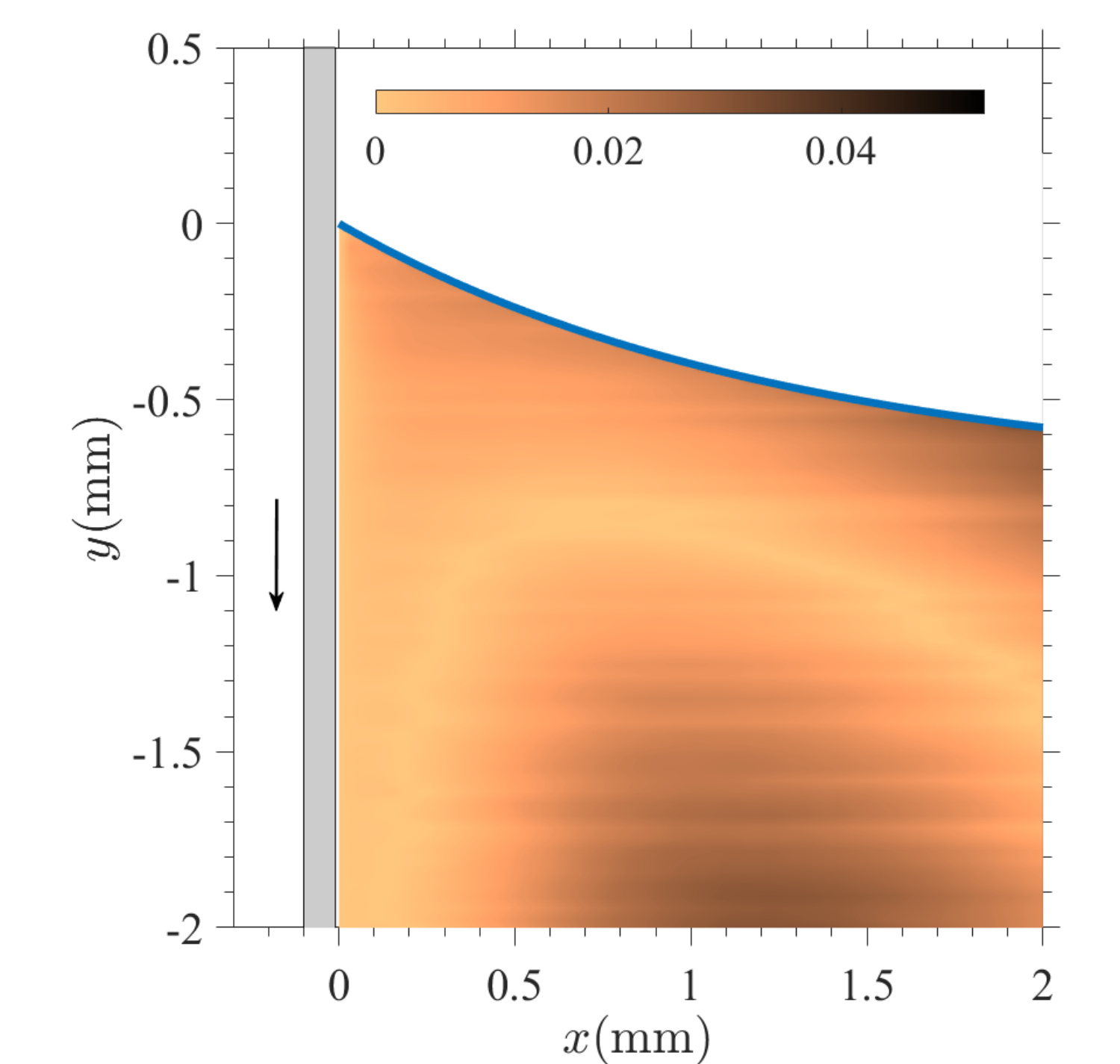}
\caption{Comparison between experimentally obtained streamfunction contours and theoretical predictions shown in terms of the magnitude of the difference in streamfunction values, $|\psi - \psi_{MWS}|$. The data corresponds to figure \ref{fig:streamfunction_500cSt}. The interface is shown with a solid blue curve.}
\label{fig:streamfunction_error}
\end{figure}


For perfect agreement between experiments and theory, we expect both the contour shapes as well as contour values to agree, and this forms a very stringent test of the theory. As discussed earlier, the theory makes predictions for the streamfunction in the bulk flow in spite of suffering from a singularity at the moving contact line. On the other hand, in the experiments, it is necessary to ensure that there is no disturbance to the flow field from external disturbances. This allows the `local' nature of the flow near a moving contact line to leave its signature even at macroscopic scales. To achieve this, utmost care was taken to ensure that remnants of the flow currents caused due to initial mixing of the seeding particles dissipated substantially before the experiments were conducted. The background disturbance, quantified in terms of a ``disturbance kinetic energy'', $(u^2+v^2)/2$, at each point in the domain, was measured before the start of the experiments and was ensured to be much smaller than $U_{\text{plate}}^2/2$. Even at the lowest speeds in our experiments, the typical value of background kinetic energy at every point in the domain was less than 0.01 of the plate kinetic energy. 

Figure \ref{fig:silicone_streamfunction_varying_cSt} shows contours of streamfunction from experiments (solid curves) and modulated wedge solution (dashed curves) for four different silicone oils, 500 cSt, 100 cSt, 20 cSt, and 10 cSt respectively. The corresponding contour levels are shown for both experiments and theory and are found to be in remarkable agreement with each other, especially in the vicinity of the contact line. Flow fields are shown in a region of 2 mm $\times$ 2 mm which is comparable to the capillary length, the scale of the outer region in Cox's model. In the highly viscous experiments, i.e. experiments involving 500 cSt and 100 cSt oils shown in figures \ref{fig:streamfunction_500cSt} and \ref{fig:streamfunction_100cSt}, excellent agreement is found between experiments and theory over the entire region of interest, whereas with 20 cSt and 10 cSt oils shown in figure \ref{fig:streamfunction_20cSt}, \ref{fig:streamfunction_10cSt}, the agreement deteriorates as one approach the outer length scale. This suggests that the length scale of the 'intermediate region' of Cox (same as MWS) reduces as viscosity is reduced. This is consistent with the fact that the extent of the `intermediate' region in Cox's model is larger when viscous effects become dominant. At these viscosity ratios, the theory predicts that the flow is of ``rolling'' type, i.e., the fluid particles at the interface advect towards the moving contact line. Below the interface, fluid particles are dragged towards the plate and undergo a sharp turn near the corner of the wedge and then turn downwards near the plate. In the experiments, it was found that the rolling pattern of the flow persisted at length scales much larger than the outer length scale, the capillary length.

To further test the difference between theory and experiments, the contour of $|\psi - \psi_{MWS}|$ is shown in figure \ref{fig:streamfunction_error}. This case corresponds to the result presented in figure \ref{fig:streamfunction_500cSt}. Excellent agreement is found between experiments and theory except at large radial distances away from the contact line. This deviation is understandable since Huh \& Scriven's theory, on which the modulated wedge solution is based, is a local theory and is only valid at distances much smaller than the `outer' length scale of the problem.

The results in figure \ref{fig:silicone_streamfunction_varying_cSt} only form a small sample, but all the experiments corresponding to the parameter points shown in figure \ref{fig:operating_regime_map} at different values of $Re$ and $Ca$ yielded similar results. To the best of our knowledge, these experiments form the first such comparison in the literature where flow in the bulk was directly compared with theory. Earlier studies were largely limited to comparing the interface shapes and/or interfacial velocities.


\subsection{Interfacial speed}\label{sec:interface_speed}
Another key result from the present study is the measurement of interfacial speed. The interfacial speed is obtained by taking the projection of the experimental velocity field obtained from PIV onto the interface, where the interface is a curve that cuts the cartesian grid containing PIV data. Convergence studies were conducted to test various interpolation schemes. Tangential velocity was also determined along curves parallel to the interface and convergence was obtained when these parallel curves approached the interface from below.

\begin{figure}
\centering
\subfigure[]{\label{fig:interfacial_speed_MWS_HS_exp}
\includegraphics[trim = 0mm 0mm 0mm 0mm, clip, angle=0,width=0.75\textwidth]{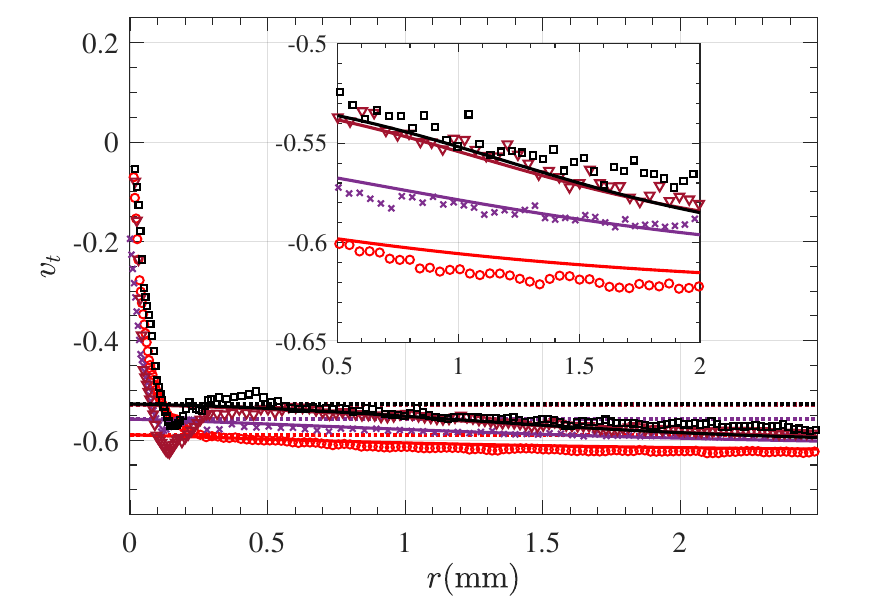}}
\subfigure[]{\label{fig:interfacial_speed_MWS_HS_exp_closeup}
\includegraphics[trim = 0mm 0mm 0mm 0mm, clip, angle=0,width=0.75\textwidth]{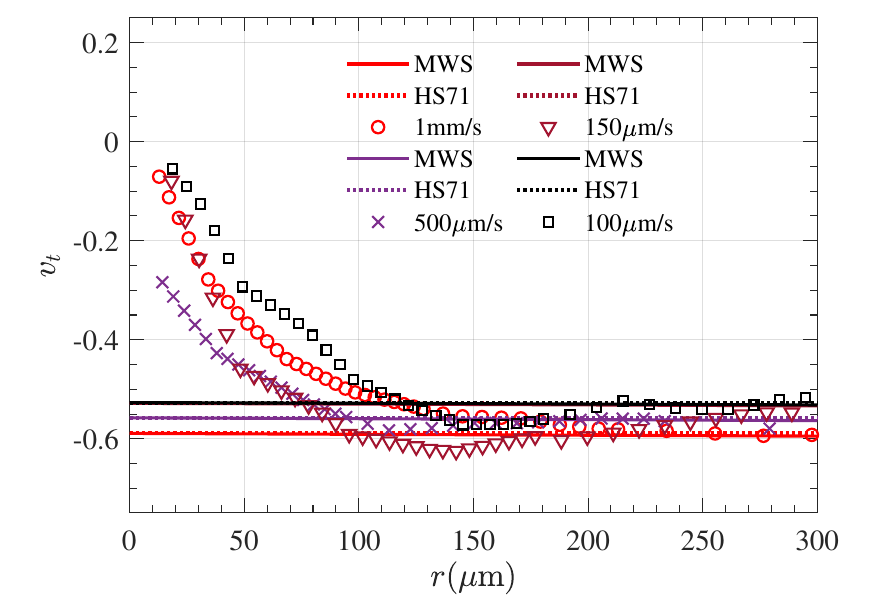}}
\caption{(a) Variation of tangential speed 
$\upsilon_t$ 
 along the curved interface where $r$ is measured from the contact line. A negative value indicates the fluid particles at the interface are moving towards the contact line consistent with the rolling motion observed in streamfunction contours. The speeds are extracted at different $Ca$ for 500 cSt silicone oil:~~ 
$\color{red}\boldsymbol{\medcirc}$: $Re = 2.61 \times 10^{-3}$,$Ca = 2.79 \times 10^{-2}$;~~~ 
$\color{violet}\boldsymbol{\bigtimes}$:$Re = 1.31 \times 10^{-3}$,$Ca = 1.40 \times 10^{-2}$;~~~ 
$\color{maroon}\triangledown$: $Re = 3.92 \times 10^{-4}$,$Ca = 4.19 \times 10^{-3}$;~~~ 
$\color{black}\boldsymbol{\square}$: $Re = 2.62 \times 10^{-4}$,$Ca = 2.79 \times 10^{-3}$. 
Interfacial speeds from experiments (different markers) are compared with the HS71 predictions (horizontal dotted lines) and  MWS theory (solid curves). The inset shows that MSW theory predicts the interfacial speed away from the contact line. (b) Close-up view of the interfacial speed near the contact line. The interface speed is non-monotonic with a slight increase at a distance of approximately 150 $\mu$m followed by a rapid reduction in speed as the contact line is approached.}
\label{fig:interfacial_speed}
\end{figure}

Figure \ref{fig:interfacial_speed} shows the variation of interfacial speed along the interface using 500 cSt silicone oil at four different speeds. Interfacial speed is plotted along the radial coordinate of the interface by fixing the origin at the contact line. The data is compared with the theoretical predictions of interfacial speed from HS71 theory and MWS given by equations \eqref{eq:v_int_HS71} and \eqref{eq:v_int_MWS} respectively. A negative value indicates that the motion is towards the moving contact line. HS71's theory predicts that the interface speed is constant, shown by dotted lines in figure \ref{fig:interfacial_speed_MWS_HS_exp}, and our experiments show fair agreement with this prediction away from the contact line, i.e. at a distance greater than $300\mu$m. A consistent but small deviation is observed between HS71 predictions and experiments which are easily accounted for by noting that the interface is curved. Without any additional fitting parameters, MWS prediction shown by solid curves, is able to capture the slight increase in speed.

At distances much closer to the contact line, typically at distances of a few hundred microns, there is a rapid deviation of the interfacial speed from classical viscous theories as shown in figure \ref{fig:interfacial_speed_MWS_HS_exp_closeup}. It has to be noted that the prediction of interfacial velocity in the classical theory of Huh \& Scriven is identical to the velocity in the intermediate region of more sophisticated theories (see figure 6 of \cite{cox1986dynamics} and figure 2 of \cite{shikhmurzaev1997moving}). These sophisticated theories resolve the singularity at the contact line by introducing addition physics in the `inner' region. One manifestation of the singularity in HS71's theory can be seen from the constant speed of approach of the fluid particles along the interface towards the contact line. This is in contrast to experiments where the interface speed rapidly decreases to nearly zero as the contact line is approached.
In summary, the experiments reveal that Huh \& Scriven's theory is valid, but only beyond a certain distance from the contact line consistent with the observation made by \cite{snoeijer2013moving}.


\section{Summary and Discussions}\label{sec:summary_discussion}
In this study, the velocity field in the vicinity of a moving contact line is determined experimentally and compared against well-known theoretical models. The nature of the flow field crucially depends on the static and dynamic contact angles, viscosity ratio, and capillary number. All these parameters are systematically varied in the current study and compared with predictions of theoretical models. Experiments are performed by vertically immersing a plate into a liquid bath at controlled speeds. Moving contact line studies can also be carried out in a number of other configurations. However, the reason behind choosing the above-mentioned configuration was to ensure that we have precise control over the speed of the contact line as well as the dynamic contact angle. Further, the two-dimensional nature of the flow in the middle of the plate allowed accurate measurement of the flow field using PIV techniques. Another important reason for choosing this configuration was to ensure that the `local' nature of the flow near the moving contact line is not affected by confinement effects which are likely to be present in the flow fields inside drops or capillary tubes. Since care was taken to ensure that no extraneous flows were present before the start of the experiment, we believe that the local flow generated due to a moving contact line can be observed at considerably larger length scales than in other flow configurations. This also allows for accurate measurement of flow fields using conventional PIV techniques. The variation of dynamic contact angle with the capillary number, shown in figure \ref{fig:Hoffmann_plot}, was found to be in excellent agreement with the classical work of \cite{hoffman1975study} which also suggests that results are configuration/geometry independent. Therefore, the obtained flow fields should be similar to flow near a moving contact line in other flow geometries.

The results are grouped into three distinct categories based on the nature of the theoretical models. In \S\ref{sec:interface_shape}, interface shapes are compared against the DRG model of \cite{dussan1991} and the generalized lubrication model of \cite{chan2020cox}. In \S\ref{sec:compare_modulated_wedge_solution}, flow fields from the experiments are compared against the \emph{modulated wedge solution (MWS)} which in turn is based on the classical work of \cite{huh1971hydrodynamic}. In \S\ref{sec:interface_speed}, interfacial speeds are compared against predictions from Huh \& Scriven's theory and the modulated wedge solution. 

All the experiments in this study involve silicone oils of varying viscosities. Since surface tension and density are nearly constant in these fluids, the capillary length, which is used as a measure of the scale of the `outer' region, is the same across all experiments. 
The Reynolds number varied from $O(10^{-4})$ to $O(1)$ and the capillary number ranged from $O(10^{-5})$ to $O(10^{-2})$.

The low $Re$ and $Ca$ in all the experiments allow us to test various types of theoretical models. The DRG model was found to be capable of determining the interface shape over a wide range of length scales except in a narrow reason near the contact line where viscous bending from the inner solution causes deviation of the interface shape. At the lowest $Ca$ values, the agreement between the DRG model and experiments was found to be nearly perfect over the entire range of spatial measurement. A test of the more recent model based on the generalisation of lubrication theory for high interface slopes yielded satisfactory results. Since the value of slip length is not known, we fitted this model for a range of slip lengths. A slip length of 14 nm was found to give the best results for all the cases. The value of this slip length is consistent with published values of slip length reported in the literature. In the case of the DRG model, the static interface shape has a large influence on the overall shape of the full dynamic interface. This is especially evident when viscous effects become small such as in the 10 cSt oil experiments. However in the case of GLM, the static interface shape is not explicitly used in the model, but the model mainly relies on viscous effects to deform the dynamic interface. It is therefore not surprising that GLM best agrees with experiments in viscous-dominated cases such as when using 500 cSt oil. A similar effect of viscosity is also observed when comparing flow fields from experiments with theory.

An ambitious test of theoretical models can be made by directly comparing the predicted velocity field in the bulk against experiments. We test this by directly comparing the streamfunction values between experiments and the modulated wedge solution, shown in figure \ref{fig:silicone_streamfunction_varying_cSt}. Excellent agreement is found for both the shape of the streamlines and the values of the streamfunction. The prediction from the modulated wedge solution is found to be within 6$\%$ deviation from the experiment as shown in figure \ref{fig:streamfunction_error}. It is worth reiterating that the HS71 or MWS solution are identical to the leading order solution in the intermediate region of \cite{cox1986dynamics}, hence, a test of HS71 is equivalent to a direct test of the intermediate solution of Cox's model. Understandably, the agreement between theory and experiment deteriorates as one moves away from the contact line since the theoretical model of \cite{huh1971hydrodynamic} is only a local theory valid in the vicinity of the contact line. In the present study, the MWS model required inputs of the local interface angle, $\beta(r)$, from experiments. The present study suggests that it is possible to predict flow fields at all length scales without any inputs from experiments, but this requires determining $\beta(r)$ from the generalised lubrication theory of \cite{chan2020cox}.

\begin{figure}
\centering
\includegraphics[trim = 0mm 0mm 0mm 0mm, clip, angle=0,width=0.40\textwidth]{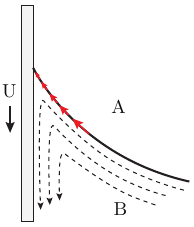}
\caption{Typical flow pattern observed in all experiments showing rolling motion in the bulk fluid (black dashed curves) consistent with figure \ref{fig:silicone_streamfunction_varying_cSt}. Fluid particles at the interface rapidly slow down as the contact line is approached consistent with figure \ref{fig:interfacial_speed}, before eventually turning down to move along the plate.}
\label{fig:Schematic_low_vel_CL}
\end{figure}


One of the most interesting findings of the present study is the measurement of the interfacial velocity. Since the fluid particles approach the contact line, the interfacial velocity is negative. Away from the contact line, the interfacial velocity is largely similar to that predicted by Huh \& Scriven's theory. This is similar to what was also observed by \cite{fuentes2005flow}, but their measurements were at much larger length scales than the present experiments. There is a small deviation in the experiments from the value predicted by Huh \& Scriven's theory. This deviation can be easily accounted for by recognizing that the interface is curved. The inset of figure \ref{fig:interfacial_speed_MWS_HS_exp} shows that accounting for the curvature of the interface in the modulated wedge solution increases the velocity marginally and is in excellent agreement with experiments. Recall that in Huh \& Scriven's theory, the interfacial velocity is independent of the radial location from the contact line. This means that fluid particles just below the interface approach the contact line at a constant speed and are required to make an instantaneous turn at the contact line, and subsequently move along the the plate. These fluid particles will be required to possess an infinite acceleration as $r \rightarrow 0$. This is yet another manifestation of Huh \& Scriven's singularity. Contrary to theoretical predictions, what we find in the experiments is that the interfacial velocity rapidly decreases to very small values as one approaches the contact line. This rapid reduction in speed to a near-zero velocity is non-monotonic. There is a small increase in the speed at a distance of approximately 150 $\mu$m from the contact line and then a rapid decrease to a very small speed as shown in figure \ref{fig:interfacial_speed_MWS_HS_exp} and \ref{fig:interfacial_speed_MWS_HS_exp_closeup}. The singularity is thus prevented by allowing the fluid to approach the contact line at smaller speeds, shown schematically in figure \ref{fig:Schematic_low_vel_CL}. These fluid particles are then expected to slip along the moving plate, but then soon adjust to the no-slip condition along the surface of the plate. To the best of our knowledge, such a behaviour of fluid particles near the contact line has not been reported in experiments before, and is completely consistent with the maxim that ``\emph{Nature abhors singularities}''.

We hope that these experimental findings will spur new theoretical developments by incorporating necessary physics in the inner and intermediate regions. Further, quantitative data about interface shape and flow fields in the viscous phase is available for download at \cite{supplementary_data} and we hope new computational models will be validated against the PIV data provided in this paper.


\vspace{2mm}
\backsection[Supplementary data]{\label{SupMat}Supplementary material is available at \\\url{https://people.iith.ac.in/hdixit/MCL_Silicone_oil_Supplementary.html}}

\backsection[Acknowledgements]{We wish to acknowledge the support of the Science and Engineering Research Board (SERB), Dept. of Science and Technology, India for funding this research through grant no. CRG/2021/007096.}


\backsection[Declaration of interests]{The authors report no conflict of interest.}

\backsection[Data availability statement]{The data that support the findings of this study are openly available on Corresponding Author's website at \url{https://people.iith.ac.in/hdixit/MCL_Silicone_oil_Supplementary.html}. See JFM's \href{https://www.cambridge.org/core/journals/journal-of-fluid-mechanics/information/journal-policies/research-transparency}{research transparency policy} for more information}

\backsection[Author ORCIDs]{Authors may include the ORCID identifers as follows.  Charul Gupta, https://orcid.org/0009-0003-5518-2999; Anjishnu Choudhury, https://orcid.org/0000-0001-9814-5483, Lakshmana D Chandrala, https://orcid.org/0000-0002-2695-9469; Harish N Dixit, https://orcid.org/0000-0003-2993-7633}

\backsection[Author contributions]{HND conceptualized and supervised the research. AC performed the initial experiments and contributed to the analysis of interface shape. CG performed most of the experiments in this study and analyzed the data with help from LDC and HND. CG, LDC and HND wrote the manuscript.}

\appendix

\bibliographystyle{jfm}
\bibliography{viscous}

\end{document}